\input harvmac
\input epsf
%\input srctex.sty
%\input xy
%\xyoption{all}
\noblackbox

%%% Paragram
\newcount\figno

\figno=0
 
\newcount\figno
\figno=0
\def\fig#1#2#3{
\par\begingroup\parindent=0pt\leftskip=1cm\rightskip=1cm\parindent=0pt
\baselineskip=11pt
\global\advance\figno by 1
\midinsert
\epsfxsize=#3
\centerline{\epsfbox{#2}}
\vskip 12pt
{\bf Fig. \the\figno:} #1\par
\endinsert\endgroup\par
}
\def\figlabel#1{\xdef#1{\the\figno}}
\def\encadremath#1{\vbox{\hrule\hbox{\vrule\kern8pt\vbox{\kern8pt
\hbox{$\displaystyle #1$}\kern8pt}
\kern8pt\vrule}\hrule}}

\def\frac#1#2{{#1\over #2}}
\def\pano{\par\noindent}

%%% special math symbols
\font\cmss=cmss10
\font\cmsss=cmss10 at 7pt

\def\rlx{\relax\leavevmode}
\def\inbar{\vrule height1.5ex width.4pt depth0pt}
\def\IC{\relax\,\hbox{$\inbar\kern-.3em{\rm C}$}}
\def\IR{\relax{\rm I\kern-.18em R}}
\def\IQ{\relax{\rm I\kern-.44em Q}}
\def\IN{\relax{\rm I\kern-.18em N}}
\def\IP{\relax{\rm I\kern-.18em P}}
\def\ZZ{\rlx\leavevmode\ifmmode\mathchoice{\hbox{\cmss Z\kern-.4em Z}}
 {\hbox{\cmss Z\kern-.4em Z}}{\lower.9pt\hbox{\cmsss Z\kern-.36em Z}}
 {\lower1.2pt\hbox{\cmsss Z\kern-.36em Z}}\else{\cmss Z\kern-.4em Z}\fi}

%%% misc.
\def\narrowplus{\kern -.04truein + \kern -.03truein}
\def\narrowminus{- \kern -.04truein}
\def\narrowminussub{\kern -.02truein - \kern -.01truein}

\def\a{\alpha}

\def\o#1{\overline{#1}}
\def\la{\langle}
\def\ra{\rangle}

%%% further macros

%%% References

\lref\jan{
T.~W.~Grimm and J.~Louis,
``The effective action of N = 1 Calabi-Yau orientifolds,''
Nucl.\ Phys.\ B {\bf 699}, 387 (2004)
[arXiv:hep-th/0403067]\semi
%%CITATION = HEP-TH 0403067;%%
H.~Jockers and J.~Louis,
``The effective action of D7-branes in N = 1 Calabi-Yau orientifolds,''
arXiv:hep-th/0409098.
%%CITATION = HEP-TH 0409098;%%
}

\lref\fluxsam{
A.~Strominger,
``Superstrings With Torsion,''
Nucl.\ Phys.\ B {\bf 274}, 253 (1986)\semi
%%CITATION = NUPHA,B274,253;%%
J.~Polchinski and A.~Strominger,
``New Vacua for Type II String Theory,''
Phys.\ Lett.\ B {\bf 388}, 736 (1996)
[arXiv:hep-th/9510227]\semi
%%CITATION = HEP-TH 9510227;%%
K.~Becker and M.~Becker,
``M-Theory on Eight-Manifolds,''
Nucl.\ Phys.\ B {\bf 477}, 155 (1996)
[arXiv:hep-th/9605053]\semi
%%CITATION = HEP-TH 9605053;%%
J.~Michelson,
``Compactifications of type IIB strings to four dimensions with  non-trivial
 classical potential,''
Nucl.\ Phys.\ B {\bf 495}, 127 (1997)
[arXiv:hep-th/9610151]\semi
%%CITATION = HEP-TH 9610151;%%
B.~R.~Greene, K.~Schalm and G.~Shiu,
``Warped compactifications in M and F theory,''
Nucl.\ Phys.\ B {\bf 584}, 480 (2000)
[arXiv:hep-th/0004103]\semi
%%CITATION = HEP-TH 0004103;%%
M.~Grana and J.~Polchinski,
``Supersymmetric three-form flux perturbations on AdS(5),''
Phys.\ Rev.\ D {\bf 63}, 026001 (2001)
[arXiv:hep-th/0009211]\semi
%%CITATION = HEP-TH 0009211;%%
G.~Curio, A.~Klemm, D.~L\"ust and S.~Theisen,
``On the vacuum structure of type II string compactifications on  Calabi-Yau
 spaces with H-fluxes,''
Nucl.\ Phys.\ B {\bf 609}, 3 (2001)
[arXiv:hep-th/0012213].
%%CITATION = HEP-TH 0012213;%%
T.~R.~Taylor and C.~Vafa,
``RR flux on Calabi-Yau and partial supersymmetry breaking,''
Phys.\ Lett.\ B {\bf 474}, 130 (2000)
[arXiv:hep-th/9912152].
%%CITATION = HEP-TH 9912152;%%
P.~Mayr,
``On supersymmetry breaking in string theory and its realization in brane
worlds,''
Nucl.\ Phys.\ B {\bf 593}, 99 (2001)
[arXiv:hep-th/0003198].
%%CITATION = HEP-TH 0003198;%%
}

\lref\deSitter{
C.~Escoda, M.~Gomez-Reino and F.~Quevedo,
``Saltatory de Sitter string vacua,''
JHEP {\bf 0311}, 065 (2003)
[arXiv:hep-th/0307160];\semi
%%CITATION = HEP-TH 0307160;%%
C.~P.~Burgess, R.~Kallosh and F.~Quevedo,
``de Sitter string vacua from supersymmetric D-terms,''
JHEP {\bf 0310}, 056 (2003)
[arXiv:hep-th/0309187].
%%CITATION = HEP-TH 0309187;%%
}

%\AcharyaKV
\lref\AcharyaKV{
B.~S.~Acharya,
``A moduli fixing mechanism in M theory,''
arXiv:hep-th/0212294 \semi
%%CITATION = HEP-TH 0212294;%%
A.~R.~Frey and J.~Polchinski,
``N = 3 warped compactifications,''
Phys.\ Rev.\ D {\bf 65}, 126009 (2002)
[arXiv:hep-th/0201029]\semi
%%CITATION = HEP-TH 0201029;%%
O.~DeWolfe and S.~B.~Giddings,
``Scales and hierarchies in warped compactifications and brane worlds,''
Phys.\ Rev.\ D {\bf 67}, 066008 (2003)
[arXiv:hep-th/0208123]\semi
%%CITATION = HEP-TH 0208123;%%
P.~K.~Tripathy and S.~P.~Trivedi,
``Compactification with flux on K3 and tori,''
JHEP {\bf 0303}, 028 (2003)
[arXiv:hep-th/0301139]\semi
%%CITATION = HEP-TH 0301139;%%
A.~Giryavets, S.~Kachru, P.~K.~Tripathy and S.~P.~Trivedi,
``Flux compactifications on Calabi-Yau threefolds,''
JHEP {\bf 0404}, 003 (2004)
[arXiv:hep-th/0312104].
%%CITATION = HEP-TH 0312104;%%
}

%\GorlichQM
\lref\GorlichQM{
L.~G\"orlich, S.~Kachru, P.~K.~Tripathy and S.~P.~Trivedi,
``Gaugino condensation and nonperturbative superpotentials in flux
compactifications,''
arXiv:hep-th/0407130.
%%CITATION = HEP-TH 0407130;%%
}

%\WittenDF
\lref\WittenDF{
E.~Witten,
``Constraints On Supersymmetry Breaking,''
Nucl.\ Phys.\ B {\bf 202}, 253 (1982).
%%CITATION = NUPHA,B202,253;%%
}

\lref\rangles{M.~Berkooz, M.~R.~Douglas and R.~G.~Leigh, {" Branes Intersecting
at Angles"}, Nucl. Phys. B {\bf 480} (1996) 265, hep-th/9606139\semi
%%CITATION = HEP-TH 9606139;%%
C.~Bachas, {" A Way to Break Supersymmetry"}, hep-th/9503030\semi
%%CITATION = HEP-TH 9503030;%%
R.~Blumenhagen, L.~G\"orlich, B.~K\"ors and D.~L\"ust,
{"Noncommutative Compactifications of Type I Strings on Tori with Magnetic
Background Flux"}, JHEP {\bf 0010} (2000) 006, hep-th/0007024\semi
%%CITATION = HEP-TH 0007024;%%
C.~Angelantonj, I.~Antoniadis, E.~Dudas, A.~Sagnotti, {" Type I
Strings on Magnetized Orbifolds and Brane Transmutation"},
Phys. Lett. B {\bf 489} (2000) 223, hep-th/0007090\semi
%%CITATION = HEP-TH 0007090;%%
G.~Aldazabal, S.~Franco, L.~E.~Ibanez, R.~Rabadan, A.~M.~Uranga,
{"Intersecting Brane Worlds"}, JHEP {\bf 0102} (2001) 047, hep-ph/0011132 \semi
%%CITATION = HEP-PH 0011132;%%
G.~Aldazabal, S.~Franco, L.~E.~Ibanez, R.~Rabadan, A.~M.~Uranga,
{" $D=4$ Chiral String Compactifications from Intersecting Branes "},
J.\ Math.\ Phys.\  {\bf 42} (2001) 3103, hep-th/0011073\semi
%%CITATION = HEP-TH 0011073;%%
L.~E.~Ibanez, F.~Marchesano and R.~Rabadan,
``Getting just the standard model at intersecting branes,''
JHEP {\bf 0111}, 002 (2001)
[arXiv:hep-th/0105155]\semi
%%CITATION = HEP-TH 0105155;%%
R.~Blumenhagen, B.~K\"ors, D.~L\"ust and T.~Ott,
``The standard model from stable intersecting brane world orbifolds,''
Nucl.\ Phys.\ B {\bf 616}, 3 (2001)
[arXiv:hep-th/0107138]\semi
%%CITATION = HEP-TH 0107138;%%
G.~Honecker and T.~Ott,
``Getting just the supersymmetric standard model at intersecting branes on the
Z(6)-orientifold,''
arXiv:hep-th/0404055.
%%CITATION = HEP-TH 0404055;%%
}

\lref\ibwrev{
A.~M.~Uranga,
``Chiral four-dimensional string compactifications with intersecting
D-branes,''
Class.\ Quant.\ Grav.\  {\bf 20}, S373 (2003)
[arXiv:hep-th/0301032]\semi
%%CITATION = HEP-TH 0301032;%%
E.~Kiritsis,
``D-branes in standard model building, gravity and cosmology,''
Fortsch.\ Phys.\  {\bf 52}, 200 (2004)
[arXiv:hep-th/0310001]\semi
%%CITATION = HEP-TH 0310001;%%
D.~L\"ust,
``Intersecting brane worlds: A path to the standard model?,''
Class.\ Quant.\ Grav.\  {\bf 21}, S1399 (2004)
[arXiv:hep-th/0401156].
%%CITATION = HEP-TH 0401156;%%
}

\lref\rbkl{
R.~Blumenhagen, B.~K\"ors and D.~L\"ust,
{" Type I Strings with $F$ and $B$-Flux"}, JHEP {\bf 0102} (2001) 030,
hep-th/0012156.
%%CITATION = HEP-TH 0012156;%%
}

\lref\bghlw{R.~Blumenhagen, F.~Gmeiner, G.~Honecker, D.~L\"ust and T.~Weigand,
{"work in progress"}.
}

\lref\rbbkl{R.~Blumenhagen, V.~Braun, B.~K\"ors and D.~L\"ust,
{" Orientifolds of K3 and Calabi-Yau Manifolds with Intersecting D-branes"},
JHEP {\bf 0207} (2002) 026, hep-th/0206038\semi
%%CITATION = HEP-TH 0206038;%%
R.~Blumenhagen, V.~Braun, B.~K\"ors and D.~L\"ust,
{" The Standard Model on the Quintic"}, hep-th/0210083.
%%CITATION = HEP-TH 0210083;%%
}

%\BlumenhagenGW
\lref\BlumenhagenGW{
R.~Blumenhagen, L.~G\"orlich and T.~Ott,
``Supersymmetric intersecting branes on the type IIA T**6/Z(4) orientifold,''
JHEP {\bf 0301}, 021 (2003)
[arXiv:hep-th/0211059].
%%CITATION = HEP-TH 0211059;%%
}

%\BlumenhagenVR
\lref\BlumenhagenVR{
R.~Blumenhagen, D.~L\"ust and T.~R.~Taylor,
``Moduli stabilization in chiral type IIB orientifold models with fluxes,''
Nucl.\ Phys.\ B{\bf 663},319 (2003)
[arXiv:hep-th/0303016]\semi
%%CITATION = HEP-TH 0303016;%%
J.~F.~G.~Cascales and A.~M.~Uranga,
``Chiral 4d N = 1 string vacua with D-branes and NSNS and RR fluxes,''
JHEP {\bf 0305}, 011 (2003)
[arXiv:hep-th/0303024]\semi
%%CITATION = HEP-TH 0303024;%%
J.~F.~G.~Cascales and A.~M.~Uranga,
``Chiral 4d string vacua with D-branes and moduli stabilization,''
arXiv:hep-th/0311250\semi
%%CITATION = HEP-TH 0311250;%%
A.~Font,
``Z(N) orientifolds with flux,''
arXiv:hep-th/0410206.
%%CITATION = HEP-TH 0410206;%%
}

%\MarchesanoXZ
\lref\MarchesanoXZ{
F.~Marchesano and G.~Shiu,
``Building MSSM flux vacua,''
arXiv:hep-th/0409132\semi
%%CITATION = HEP-TH 0409132;%%
F.~Marchesano and G.~Shiu,
``MSSM vacua from flux compactifications,''
arXiv:hep-th/0408059\semi
%%CITATION = HEP-TH 0408059;%%
F.~Marchesano, G.~Shiu and L.~T.~Wang,
``Model Building and Phenomenology of Flux-Induced Supersymmetry Breaking on
D3-branes,''
arXiv:hep-th/0411080.
%%CITATION = HEP-TH 0411080;%%
}

%\CveticXX
\lref\CveticXX{
M.~Cvetic and T.~Liu,
``Supersymmetric Standard Models, Flux Compactification and Moduli
Stabilization,'' 
arXiv:hep-th/0409032.
%%CITATION = HEP-TH 0409032;%%
}

\lref\rcvetica{M.~Cvetic, G.~Shiu and  A.~M.~Uranga,  {" Three-Family
Supersymmetric Standard-like Models from Intersecting Brane Worlds"}
Phys. Rev. Lett. {\bf 87} (2001) 201801,  hep-th/0107143\semi
%%CITATION = HEP-TH 0107143;%%
M.~Cvetic, G.~Shiu and  A.~M.~Uranga,  {"
Chiral Four-Dimensional N=1 Supersymmetric Type IIA Orientifolds from
Intersecting D6-Branes"}, Nucl. Phys. B {\bf 615} (2001) 3, hep-th/0107166\semi
%%CITATION = HEP-TH 0107166;%%
M.~Berkooz and R.~G.~Leigh,
``A D = 4 N = 1 orbifold of type I strings,''
Nucl.\ Phys.\ B {\bf 483} (1997) 187, hep-th/9605049\semi
%%CITATION = HEP-TH 9605049;%%
S.~F\"orste, G.~Honecker and R.~Schreyer,
``Supersymmetric Z(N) x Z(M) orientifolds in 4D with D-branes at angles,''
Nucl. Phys. B {\bf 593} (2001) 127, hep-th/0008250.
%%CITATION = HEP-TH 0008250;%%
}

%\DasguptaSS
\lref\DasguptaSS{
K.~Dasgupta, G.~Rajesh and S.~Sethi,
{" M theory, orientifolds and G-flux"},
JHEP {\bf 9908}, 023 (1999), hep-th/9908088.
%%CITATION = HEP-TH 9908088;%%
}

\lref\radd{N.~Arkani-Hamed, S.~Dimopoulos, and G.~Dvali, {" The Hierarchy
Problem and New Dimensions at a Millimeter"}, Phys. Lett. B {\bf 429} (1998)
263, hep-ph/9803315\semi
%%CITATION = HEP-PH 9803315;%%
I.~Antoniadis, N.~Arkani-Hamed, S.~Dimopoulos, and G.~Dvali, {"
New Dimensions at a Millimeter to a Fermi and Superstrings at a TeV"},
Phys. Lett. B {\bf 436} (1998) 257, hep-ph/9804398.
%%CITATION = HEP-PH 9804398;%%
}

\lref\ArkaniHamedFB{
N.~Arkani-Hamed and S.~Dimopoulos,
``Supersymmetric unification without low energy supersymmetry and signatures
for fine-tuning at the LHC,''
arXiv:hep-th/0405159.
%%CITATION = HEP-TH 0405159;%%
}

%\AntoniadisDT
\lref\AntoniadisDT{
I.~Antoniadis and S.~Dimopoulos,
``Splitting Supersymmetry in String Theory,''
arXiv:hep-th/0411032\semi
%%CITATION = HEP-TH 0411032;%%
C.~Kokorelis,
``Standard models and split supersymmetry from intersecting brane orbifolds,''
arXiv:hep-th/0406258.
%%CITATION = HEP-TH 0406258;%%
}

\lref\rgimpol{
E.~G.~Gimon and J.~Polchinski, ``Consistency Conditions
for Orientifolds and D-Manifolds'', Phys.\ Rev.\ {\bf D54} (1996) 1667,
hep-th/9601038.
%%CITATION = HEP-TH 9601038;%%
}

% Statistics

\lref\wgkt{
O.~De Wolfe, A.~Giryavets, S.~Kachru and W.~Taylor,
``Enumerating Flux Vacua with Enhanced Symmetries,''
arXiv:hep-th/0411061.
%%CITATION = HEP-TH 0411061;%%
}

%\DouglasZG
\lref\DouglasZG{
M.~R.~Douglas,
``Basic results in vacuum statistics,''
arXiv:hep-th/0409207.
%%CITATION = HEP-TH 0409207;%%
}

%\DouglasKC
\lref\DouglasKC{
M.~R.~Douglas, B.~Shiffman and S.~Zelditch,
``Critical points and supersymmetric vacua, II: Asymptotics and extremal
metrics,''
arXiv:math.cv/0406089.
%%CITATION = MATH-CV 0406089;%%
}

%\DouglasQG
\lref\DouglasQG{
M.~R.~Douglas,
``Statistical analysis of the supersymmetry breaking scale,''
arXiv:hep-th/0405279.
%%CITATION = HEP-TH 0405279;%%
}

%\DenefDM
\lref\DenefDM{
F.~Denef, M.~R.~Douglas and B.~Florea,
``Building a better racetrack,''
JHEP {\bf 0406}, 034 (2004)
[arXiv:hep-th/0404257].
%%CITATION = HEP-TH 0404257;%%
}

%\DenefZE
\lref\DenefZE{
F.~Denef and M.~R.~Douglas,
``Distributions of flux vacua,''
JHEP {\bf 0405}, 072 (2004)
[arXiv:hep-th/0404116].
%%CITATION = HEP-TH 0404116;%%
}

%\DouglasKP
\lref\DouglasKP{
M.~R.~Douglas,
``Statistics of string vacua,''
arXiv:hep-ph/0401004.
%%CITATION = HEP-PH 0401004;%%
}

%\AshokGK
\lref\AshokGK{
S.~Ashok and M.~R.~Douglas,
``Counting flux vacua,''
JHEP {\bf 0401}, 060 (2004)
[arXiv:hep-th/0307049].
%%CITATION = HEP-TH 0307049;%%
}

%\DouglasUM
\lref\DouglasUM{
M.~R.~Douglas,
``The statistics of string / M theory vacua,''
JHEP {\bf 0305}, 046 (2003)
[arXiv:hep-th/0303194].
%%CITATION = HEP-TH 0303194;%%
}

%\DineCT
\lref\DineCT{
M.~Dine,
``Supersymmetry, naturalness and the landscape,''
arXiv:hep-th/0410201.
%%CITATION = HEP-TH 0410201;%%
}

%\DineIS
\lref\DineIS{
M.~Dine, E.~Gorbatov and S.~Thomas,
``Low energy supersymmetry from the landscape,''
arXiv:hep-th/0407043.
%%CITATION = HEP-TH 0407043;%%
}

%\SusskindUV
\lref\SusskindUV{
L.~Susskind,
``Supersymmetry breaking in the anthropic landscape,''
arXiv:hep-th/0405189.
%%CITATION = HEP-TH 0405189;%%
}

%\WeinbergCP
\lref\WeinbergCP{
S.~Weinberg,
``The Cosmological Constant Problem,''
Rev.\ Mod.\ Phys.\  {\bf 61}, 1 (1989).
%%CITATION = RMPHA,61,1;%%
}

%\SusskindKW
\lref\SusskindKW{
L.~Susskind,
``The anthropic landscape of string theory,''
arXiv:hep-th/0302219.
%%CITATION = HEP-TH 0302219;%%
}

%\ConlonDS
\lref\ConlonDS{
A.~Misra and A.~Nanda,
``Flux vacua statistics for two-parameter Calabi-Yau's,''
arXiv:hep-th/0407252\semi
%%CITATION = HEP-TH 0407252;%%
J.~P.~Conlon and F.~Quevedo,
``On the explicit construction and statistics of Calabi-Yau flux vacua,''
arXiv:hep-th/0409215.
%%CITATION = HEP-TH 0409215;%%
}

%\GiryavetsZR
\lref\GiryavetsZR{
A.~Giryavets, S.~Kachru and P.~K.~Tripathy,
``On the taxonomy of flux vacua,''
JHEP {\bf 0408}, 002 (2004)
[arXiv:hep-th/0404243].
%%CITATION = HEP-TH 0404243;%%
}

%\KachruAW
\lref\KachruAW{
S.~Kachru, R.~Kallosh, A.~Linde and S.~P.~Trivedi,
``De Sitter vacua in string theory,''
Phys.\ Rev.\ D {\bf 68}, 046005 (2003)
[arXiv:hep-th/0301240].
%%CITATION = HEP-TH 0301240;%%
}

%\KachruHE
\lref\KachruHE{
S.~Kachru, M.~B.~Schulz and S.~Trivedi,
``Moduli stabilization from fluxes in a simple IIB orientifold,''
JHEP {\bf 0310}, 007 (2003)
[arXiv:hep-th/0201028].
%%CITATION = HEP-TH 0201028;%%
}

%\GiddingsYU
\lref\GiddingsYU{
S.~B.~Giddings, S.~Kachru and J.~Polchinski,
``Hierarchies from fluxes in string compactifications,''
Phys.\ Rev.\ D {\bf 66}, 106006 (2002)
[arXiv:hep-th/0105097].
%%CITATION = HEP-TH 0105097;%%
}

%\KumarPV
\lref\KumarPV{
J.~Kumar and J.~D.~Wells,
``Landscape cartography: A coarse survey of gauge group rank and stabilization
%of the proton,''
arXiv:hep-th/0409218.
%%CITATION = HEP-TH 0409218;%%
}

%\BoussoXA
\lref\BoussoXA{
R.~Bousso and J.~Polchinski,
``Quantization of four-form fluxes and dynamical neutralization of the
cosmological constant,''
JHEP {\bf 0006}, 006 (2000)
[arXiv:hep-th/0004134].
%%CITATION = HEP-TH 0004134;%%
}

%\LustDN
\lref\LustDN{
P.~G.~Camara, L.~E.~Ibanez and A.~M.~Uranga,
``Flux-induced SUSY-breaking soft terms,''
Nucl.\ Phys.\ B {\bf 689}, 195 (2004)
[arXiv:hep-th/0311241]\semi
%%CITATION = HEP-TH 0311241;%%
M.~Grana, T.~W.~Grimm, H.~Jockers and J.~Louis,
``Soft supersymmetry breaking in Calabi-Yau orientifolds with D-branes and
fluxes,''
Nucl.\ Phys.\ B {\bf 690}, 21 (2004)
[arXiv:hep-th/0312232]\semi
%%CITATION = HEP-TH 0312232;%%
D.~L\"ust, S.~Reffert and S.~Stieberger,
``Flux-induced soft supersymmetry breaking in chiral type IIb orientifolds with
D3/D7-branes,''
arXiv:hep-th/0406092\semi
%%CITATION = HEP-TH 0406092;%%
P.~G.~Camara, L.~E.~Ibanez and A.~M.~Uranga,
``Flux-induced SUSY-breaking soft terms on D7-D3 brane systems,''
arXiv:hep-th/0408036\semi
%%CITATION = HEP-TH 0408036;%%
L.~E.~Ibanez,
``The fluxed MSSM,''
arXiv:hep-ph/0408064\semi
%%CITATION = HEP-PH 0408064;%%
D.~L\"ust, S.~Reffert and S.~Stieberger,
``MSSM with soft SUSY breaking terms from D7-branes with fluxes,''
arXiv:hep-th/0410074.
%%CITATION = HEP-TH 0410074;%%
}

\lref\wong{
R.~Wong, "Asymptotic Approximations of Integrals", Academic Press,
New York (1989). }

\lref\andrews{
G.E.~Andrews, "The Theory of Partitions", Encyclopedia of Mathematics and its
Applications", Vol.2, Addison Wesley, (1976). }

%\KreuzerXY
\lref\KreuzerXY{
M.~Kreuzer and H.~Skarke,
``Complete classification of reflexive polyhedra in four dimensions,''
Adv.\ Theor.\ Math.\ Phys.\  {\bf 4}, 1209 (2002)
[arXiv:hep-th/0002240].
%%CITATION = HEP-TH 0002240;%%
}

\lref\gepner{
R.~Blumenhagen and A.~Wisskirchen,
``Spectra of 4D, N = 1 type I string vacua on non-toroidal CY threefolds,''
Phys.\ Lett.\ B {\bf 438}, 52 (1998)
[arXiv:hep-th/9806131] \semi
%%CITATION = HEP-TH 9806131;%%
G.~Aldazabal, E.~C.~Andres, M.~Leston and C.~Nunez,
``Type IIB orientifolds on Gepner points,''
JHEP {\bf 0309}, 067 (2003)
[arXiv:hep-th/0307183]\semi
%%CITATION = HEP-TH 0307183;%%
R.~Blumenhagen,
``Supersymmetric orientifolds of Gepner models,''
JHEP {\bf 0311}, 055 (2003)
[arXiv:hep-th/0310244]\semi
%%CITATION = HEP-TH 0310244;%%
I.~Brunner, K.~Hori, K.~Hosomichi and J.~Walcher,
``Orientifolds of Gepner models,''
arXiv:hep-th/0401137\semi
%%CITATION = HEP-TH 0401137;%%
R.~Blumenhagen and T.~Weigand,
``Chiral supersymmetric Gepner model orientifolds,''
JHEP {\bf 0402}, 041 (2004)
[arXiv:hep-th/0401148] \semi
%%CITATION = HEP-TH 0401148;%%
R.~Blumenhagen and T.~Weigand,
``A note on partition functions of Gepner model orientifolds,''
Phys.\ Lett.\ B {\bf 591}, 161 (2004)
[arXiv:hep-th/0403299]\semi
%%CITATION = HEP-TH 0403299;%%
G.~Aldazabal, E.~C.~Andres and J.~E.~Juknevich,
``Particle models from orientifolds at Gepner-orbifold points,''
JHEP {\bf 0405}, 054 (2004)
[arXiv:hep-th/0403262].
%%CITATION = HEP-TH 0403262;%%
}

\lref\huiszoon{
T.~P.~T.~Dijkstra, L.~R.~Huiszoon and A.~N.~Schellekens,
``Chiral supersymmetric standard model spectra from orientifolds of Gepner
 models,''
arXiv:hep-th/0403196\semi
%%CITATION = HEP-TH 0403196;%%
T.~P.~T.~Dijkstra, L.~R.~Huiszoon and A.~N.~Schellekens,
``Supersymmetric Standard Model Spectra from RCFT orientifolds,''
arXiv:hep-th/0411129.
%%CITATION = HEP-TH 0411129;%%
}

\lref\cvetic{
M.~Cvetic, I.~Papadimitriou and G.~Shiu,
``Supersymmetric three family SU(5) grand unified models from type IIA
orientifolds with intersecting D6-branes,''
Nucl.\ Phys.\ B {\bf 659}, 193 (2003)
[Erratum-ibid.\ B {\bf 696}, 298 (2004)]
[arXiv:hep-th/0212177]\semi
%%CITATION = HEP-TH 0212177;%%
M.~Cvetic and I.~Papadimitriou,
``More supersymmetric standard-like models from intersecting D6-branes on  type
IIA orientifolds,''
Phys.\ Rev.\ D {\bf 67}, 126006 (2003)
[arXiv:hep-th/0303197]\semi
%%CITATION = HEP-TH 0303197;%%
M.~Cvetic, T.~Li and T.~Liu,
``Supersymmetric Pati-Salam models from intersecting D6-branes: A road to the
standard model,''
Nucl.\ Phys.\ B {\bf 698}, 163 (2004)
[arXiv:hep-th/0403061]\semi
M.~Cvetic, P.~Langacker, T.~j.~Li and T.~Liu,
``D6-brane splitting on type IIA orientifolds,''
arXiv:hep-th/0407178.
%%CITATION = HEP-TH 0407178;%%
}

\lref\BrownDD{
J.~D.~Brown and C.~Teitelboim,
``Dynamical Neutralization Of The Cosmological Constant,''
Phys.\ Lett.\ B {\bf 195}, 177 (1987)\semi
%%CITATION = PHLTA,B195,177;%%
J.~D.~Brown and C.~Teitelboim,
``Neutralization Of The Cosmological Constant By Membrane Creation,''
Nucl.\ Phys.\ B {\bf 297}, 787 (1988).
%%CITATION = NUPHA,B297,787;%%
}

\lref\yuka{
D.~Cremades, L.~E.~Ibanez and F.~Marchesano,
``Yukawa couplings in intersecting D-brane models,''
JHEP {\bf 0307}, 038 (2003)
[arXiv:hep-th/0302105]\semi
%%CITATION = HEP-TH 0302105;%%
M.~Cvetic and I.~Papadimitriou,
``Conformal field theory couplings for intersecting D-branes on
%orientifolds,''
Phys.\ Rev.\ D {\bf 68}, 046001 (2003)
[Erratum-ibid.\ D {\bf 70}, 029903 (2004)]
[arXiv:hep-th/0303083]\semi
%%CITATION = HEP-TH 0303083;%%
S.~A.~Abel and A.~W.~Owen,
``Interactions in intersecting brane models,''
Nucl.\ Phys.\ B {\bf 663}, 197 (2003)
[arXiv:hep-th/0303124]\semi
%%CITATION = HEP-TH 0303124;%%
D.~Cremades, L.~E.~Ibanez and F.~Marchesano,
``Computing Yukawa couplings from magnetized extra dimensions,''
JHEP {\bf 0405}, 079 (2004)
[arXiv:hep-th/0404229].
%%CITATION = HEP-TH 0404229;%%
}

%%% Title page
\Title{\vbox{
 \hbox{MPP-2004-147}
\hbox{LMU-TPS 04/13}
 \hbox{hep-th/0411173}}}
 %\vskip-1cm
{\vbox{\centerline{The Statistics of Supersymmetric D-brane Models}
}}
\centerline{Ralph Blumenhagen$^a$, Florian Gmeiner$^a$, Gabriele Honecker$^a$,}
\centerline{ Dieter L\"ust$^{a,b}$  and  Timo Weigand$^a$ }
\bigskip\medskip
\centerline{{\it  $^a$Max-Planck Institut f\"ur Physik, 
F\"ohringer Ring 6, 80805 M\"unchen, Germany }}
\centerline{\tt e-mail:
blumenha, flo, gabriele, luest, weigand@mppmu.mpg.de}

\bigskip\medskip
\centerline{{\it  $^b$Department f\"ur Physik, Ludwig-Maximilians-Universit\"at M\"unchen}}
\centerline{{\it Theresienstra{\ss}e 37, 80333 M\"unchen, Germany }}
\centerline{\tt e-mail:
luest@theorie.physik.uni-muenchen.de}

\bigskip
\bigskip

\centerline{\bf Abstract}
\noindent
We investigate the statistics of the phenomenologically important
D-brane sector of string compactifications.
In particular for the class of intersecting D-brane models, we generalise
methods known from number theory 
to determine the asymptotic statistical distribution of solutions
to the tadpole cancellation conditions.
Our approach  allows us to compute the statistical distribution
of gauge theoretic observables like the rank of the gauge group,
the number of chiral generations or the probability of an $SU(N)$ gauge factor.
Concretely,  
we study the statistics of  intersecting branes on  $T^2$ and  $T^4/\ZZ_2$ and
$T^6/\ZZ_2\times \ZZ_2$ orientifolds. 
Intriguingly, we find a statistical correlation between the rank
of the gauge group and the number of chiral generations. 
Finally, we  combine the statistics of the gauge theory sector
with the statistics of the flux sector and study how
 distributions of gauge theoretic  quantities are affected.

%\medskip

\Date{11/2004}
%%% text
\newsec{Introduction}

For many years,  string theorists have studied various kinds 
of string compactifications with the main motivation 
to find a model one day which
resembles the Standard Model of particle physics,
which we know is a good effective description of nature
around the weak scale. Of course, it is very
compelling that string theory in some sense predicts
(or at least contains)  gravity and at the same time also gauge theory, but
it turned out to be a very difficult problem to
be more precise about the predictions 
for  the gauge theory sector at low energies. 
There were simply too many possible candidates
which at best resembled the Standard Model in certain
aspects but not in its full glory. However, one
could also say that if we had 
constructed the Standard Model already in  a few
attempts, we could have simply been very lucky
or more probably would have just found one of a 
vast number of stringy realizations of the Standard Model.

Recall that  in the eighties the weakly coupled 
heterotic string was thought to be the best candidate
to deliver a realistic model. 
Various (supersymmetric) constructions, like
orbifolds, bosonic lattices,
free fermion constructions, Calabi-Yau
manifolds in toric varieties, Landau-Ginzburg and Gepner 
models etc.
were studied and partly classified in a large number of publications.
Even today it is not clear whether there
exist finitely or infinitely many   heterotic string backgrounds.

With the advent of D-branes in the mid nineties, new classes
of   phenomenologically attractive string compactifications were found where
the gauge theory is realized by the open strings ending
on D-branes. Different constructions were proposed
like orientifolds with branes on singularities or
intersecting D-branes. It was also realized that
the string scale does not necessarily have to
be close to the Planck scale, as in D-brane constructions
low string scale models can exist and would lead to a completely
new phenomenology in the TeV range \refs{\radd}. In this case,
supersymmetry would no longer be necessary 
for solving the gauge hierarchy problem.

In addition, with the extension of the 
relevant dimensions
to eleven and even twelve, new geometric compactifications
were possible like M-theory on $G_2$ manifolds or 
F-theory on Calabi-Yau fourfolds. 
Since among these different classes of string models  various
$S$ and $T$ dualities operate, not all constructions
were considered to be  independent, but rather provide  descriptions in different
regimes of the M-theory moduli space.
It was clear by then that there exist very many string vacua, but
nevertheless people hoped that maybe one could find a realistic model
out of these, say,  $10^{10}$ vacua \foot{Numbers of this order appeared for 
instance in the classification of CY fourfolds \KreuzerXY.}.

Moreover, with the study of flux compactifications 
\refs{\fluxsam,\DasguptaSS,\GiddingsYU,\KachruHE,\AcharyaKV}
some additional progress was made:
First, flux compactifications
allowed to solve some of the problems the former
purely geometric models notoriously had. In particular, certain fluxes
induce an effective potential that still possesses
supersymmetric minima, which allows to freeze (some of)
the moduli generically appearing in string theory. 
Second, it was also possible 
to break supersymmetry in a controlled way by turning on additional
internal flux components. Finally,
by taking also some non-perturbative
effects into account, for the first time
strong evidence was given  that non-supersymmetric meta-stable
de-Sitter vacua do exist in string theory 
\refs{\KachruAW,\deSitter,\DenefDM,\GorlichQM}.
Concerning the classification of string vacua, these
flux compactifications were also quite surprising, as 
it turned out that there exist astronomical numbers of 
minima of the flux induced potential. One is now not talking
about $10^{10}$ vacua but about $10^{500}$  
different vacua in the string theory landscape.

M.R. Douglas emphasized that the search for {\it the} Standard Model
like vacuum does not really make sense in view of  this
huge number of string models \DouglasUM. Given the precisions with
which the Standard Model parameters are measured there
probably exists still a large amount of stringy realizations
consistent with these precisions.
Of course, finding one of these string models would still be a big success from
the phenomenological point of view,
as once the background is fixed there probably
exist fewer parameters than in the Standard Model.
%\foot{However,  
%finding it probably needs more than the effort
%of a single physicist with a desk-top in his office.
%Even for the computations performed in this paper we 
%were using a cluster of $\sim$60 computers.}

Contrarily, M.R. Douglas proposed a different complementary
approach to the
string vacuum problem. Given the huge numbers which are
occurring, one had better try a statistical approach.
Instead of computing phenomenologically
important quantities for each model separately, we
ought to compute expectation values or distributions of these quantities
in the ensemble of string vacua. 
Based on  this approach, it was proposed  that some of the  fine
tuning problems like for the cosmological constant or for the
weak scale can find a purely statistical explanation on 
the string landscape.
If the amount $\epsilon$ of  fine tuning is larger  than the density of the supposedly
uniformly distributed string models, $\epsilon>1/N$, 
there can still be a sufficient  number of string vacua giving rise
to  the right value \refs{\BrownDD,\BoussoXA,\DouglasQG}.
Of course, questions like why nature has chosen a 
specific, maybe even unlikely value ask for a vacuum selection
principle and are beyond the statistical
approach. 
Some authors have suggested that  eventually the weak anthropic principle
might be the only clue  for ``explaining'' the actual
values of certain physical constants \refs{\WeinbergCP,\SusskindKW}. 
Whether this is the case, we don't know, but we think it
is not yet time to give up searching for  other explanations.

If also the second fine tuning problem, the gauge hierarchy problem, 
is merely solved statistically, then
there would be no need for low energy supersymmetry breaking.
There is still an ongoing debate of whether the string landscape
favours a low or a high supersymmetry breaking scale \refs{\SusskindUV\DineIS\DouglasQG-\DineCT}, 
but inspired by these string landscape ideas
already new phenomenological models with high scale
supersymmetry breaking have been proposed \ArkaniHamedFB.
It was also suggested that branes with gauge fluxes respectively
intersecting branes might provide a stringy realization
of these so-called split supersymmetry  models \AntoniadisDT. 
On the other hand, if one is still insisting in low scale supersymmetry
breaking within the observable sector of the supersymmetric Standard Model,
flux compactifications allow for an explicit computation of 
the soft supersymmetry breaking parameters (gaugino masses, squark and slepton masses
etc.) in semi-realistic models with D3- and/or D7-branes \LustDN.

At the moment we do not know whether string theory is
realized in nature.  
Without having a proof that string theory
is in a mathematical sense the  unique  quantum theory of gravity,
we can only hope to find support for string theory
experimentally. Progress in this direction
is as we know hampered  by the huge vacuum degeneracy.

It was proposed that performing statistics of the string theory
landscape could improve our understanding of the relevance
of string theory in nature. 
Indeed, one could imagine that two different physical
observables are statistically excluded 
so that effectively  they are never realized together
on the string theory landscape. Then this
would practically falsify string theory.
On the other hand one could imagine that by
computing statistical correlations of physical observables on the
string theory landscape we find more and more
support that our universe satisfies these
correlations. Of course this would not be
a proof of string theory but we could hope
to get more and more support from such considerations.

However, there exists also the possibility of the depressing 
outcome that the statistical analysis does
not show any strong correlations among various quantities
so that it does not lead to any strong statements
at all. In this scenario, all values of physical quantities are more or less
uniformly realized and everything is equally likely.
On the one hand, such uniform distributions
are an advantage for solving some of the fine tuning
problems and allow one to argue that Standard-like
models should exist on the landscape, on the other hand they are 
rather boring, as they do not yield any at least
statistical prediction.
Therefore, it should be one of the major goals
of a statistical analysis to reveal statistical
correlations among physically relevant quantities. 

So far the main and most successful work on the statistical approach deals
with the flux induced F-terms and how they
freeze the complex structure moduli. In fact, in \AshokGK\
a formula was derived for counting the index 
of vacua of the flux induced superpotential.
This formula was successfully tested in \refs{\GiryavetsZR,\ConlonDS} 
for two different Calabi-Yau manifolds with one respectively two K\"ahler 
moduli.
Phenomenologically it is of course very important
to also study the statistics of the gauge theory 
sector of string compactifications. Some general
comments and estimates were made in the original work \DouglasUM.

The aim of this paper is to develop more refined methods 
to work out the statistical distributions of physically
interesting quantities in the gauge theory sector
of string vacua.
In the original Type IIB flux models \refs{\GiddingsYU,\KachruHE} the only D-branes 
were parallel D3-branes,
on which no interesting physics like chirality can occur.
It was then proposed to augment this set-up by more
general branes, like branes on singularities or
branes with non-trivial 2-form  gauge fluxes (intersecting branes
in the T-dual picture). 
It is known that these intersecting D-brane models are constrained
by only  fairly
moderate consistency conditions 
\refs{\rangles,\rcvetica,\ibwrev}
which still allow for a large number of solutions.
This makes them  good candidates for a statistical approach.
Note that the statistics of D3-branes was discussed 
in \KumarPV.

In this paper, given a closed string background, 
we will present a method to count classically different
solutions to the tadpole cancellation conditions. 
As it stands, this is a purely number theoretical problem.
It turns out that the counting of unordered partitions is very similar
to this task. It is precisely the saddle point
method working  quite well in the former  case which can
be successfully  carried over to our problem.
It allows us to  compute the statistical distribution
of various gauge theoretic quantities and to reveal
statistical correlations among quantities which
from the gauge theory point of view  
are completely unrelated. Therefore, this can be
considered as a purely string theoretical statistical
effect and, of course, can be traced back to
the fact that string theory is a more constrained system
than gauge theory.

This paper is organised as follows. In section 2 we 
formulate the problem of counting supersymmetric solutions to
the tadpole cancellation conditions arising in intersecting
D-brane models of the Type IIA string theory.
Note that via T-duality these models are related to 
Type IIB orientifolds with even dimensional branes
with non-trivial gauge fluxes turned on.
Therefore, as worked out in 
\refs{\BlumenhagenVR,\MarchesanoXZ,\CveticXX}, 
we can combine
these models with the Type IIB flux compactifications
of \refs{\GiddingsYU,\KachruHE}. 
In a first approximation,  
the brane and the flux sector are quite independent,
the latter giving rise to F-term potentials with 
the former leading to D-term potentials.
The only effect of the flux is a contribution to
the R-R 4-form tadpole cancellation condition.

In section 3, as a toy model we study the analogous
eight dimensional problem where we have only
compactified on a two-dimensional torus.
This model is sufficiently simple, but
still close enough to our actual problem to
allow for a clear  explanation of our method.
In addition, for this simple model
we also perform a brute force computer search
for all solutions to the tadpole cancellation
condition.
Note that the semi-analytic saddle point method 
we will propose
is much more effective and far less time consuming
than this brute force numerical classification.

In section 4 we will generalise the method
to six-dimensional models on $T^4/\ZZ_2$ \rgimpol, where new
aspects like chirality and complex structure
dependence arise. We will derive the statistical
distribution for various quantities of interest
like the total number of solutions, the probability of
an $SU(M)$ gauge factor, the distribution
of the rank of the gauge group and the distribution
of the number of families. These 6D results are
again supported by a numerical computer search
for all solutions. 
Intriguingly, we find
a correlation between the rank of the gauge
group and the number of families.

In section 5 we come to the physically most interesting 
example and compute
the same distribution for the case of intersecting branes
on the $T^6/\ZZ_2\times \ZZ_2$ orientifold \refs{\rcvetica,\cvetic}.
It turns out that this case is even more complicated
than the 6D case, mainly because supersymmetry also
allows for negative contributions to the tadpole
cancellation conditions. We will show that
the number of solutions to these
conditions for fixed orientifold and 3-form flux 
contributions is  finite.
Surprisingly,  all three different models in 8D, 6D and 4D 
qualitatively show the same behaviour, so
that one can suspect that the statistics
of intersecting D-brane models in general is already
captured by our concrete examples.
For this case, some of the computations we did in 8D
and 6D are already very time consuming. Therefore,  
we restrict ourselves in this paper to a selection of those 
statistical distributions  which could  be computed  in reasonable time,
while postponing a more thorough investigation to a future publication
\bghlw.

In section 6 we combine the gauge theory sector distributions
computed in section 5 with the statistics of flux compactifications
and analyze how taking both effects into account changes
the statistics. In particular, we focus on the dependence
of the distribution of the rank of the gauge group on
the number of 3-cycles where one can turn on
three-form fluxes. 

Finally, in section 8 we give our  conclusions and an outlook of what
kind of statistical questions one might want to approach
next.
We add three appendices, in the first two of which we derive
the conditions for coprime wrapping numbers in 6D and 4D.
The third appendix briefly summarises the main technical tools we are
using. They are expected to be applicable to any problem
of the same sort. Whenever  one wants to analyse the statistics
of solutions to integer valued constraints, which are known
to have a large number of solutions, a variant of
our number theoretic formulas should work.

\newsec{Counting solutions to tadpole cancellation conditions}

In this section we formulate the problem of counting respectively computing various
statistical distributions of phenomenologically relevant observables 
in the set-up of stringy supersymmetric intersecting D-brane models. 
Even though we will also discuss eight and six-dimensional models,
let us explain the problem for the more relevant four-dimensional
case. 

\subsec{The statistical problem for general Type II orientifolds}

For concreteness we start with a Type IIB orientifold flux compactification, where
we also allow all fluxes to vanish. The F-term scalar potential generated by the fluxes
generically freezes 
the complex structure moduli and the dilaton. 
The techniques developed in \refs{\DouglasUM,\AshokGK,\DouglasKP,\DenefZE,\DouglasKC
\GiryavetsZR,\ConlonDS}
can be used  to perform
a statistical analysis of the ensemble of these
flux vacua, which for instance allows one to address questions
about the distribution of the cosmological constant and the
scale of supersymmetry breaking.

In addition, the tadpole cancellation conditions
also require the presence of D-branes, which from the phenomenological
point of view are at least equally important, as they provide the
stringy realization of the particle physics with all its interactions.
It has been pointed out that to get interesting gauge theories with chiral
matter content, one has to introduce more general branes than
simply D3-branes. Both the possibility of
branes on singularities and branes with non-trivial gauge fluxes
have been proposed \refs{\BlumenhagenVR,\MarchesanoXZ}.

If one neglects the back-reaction of the
fluxes on the background, i.e. for models at large radii with diluted 
fluxes, the D-brane sector is more or less independent
of the flux sector. The only effect is the contribution of
the fluxes to the R-R 4-form tadpole
cancellation condition.  

We would like to perform a statistical analysis for the D-brane sector
as well, where just for convenience we work from now on in the 
T-dual (mirror symmetric)
framework of intersecting D-branes.
Therefore we consider  a 
Type IIA string compactification on some  Calabi-Yau manifold, $M$, 
and  divide the Type IIA model by the discrete symmetry $\Omega \o\sigma$.
Here $\Omega$ denotes the world-sheet parity transformation and
$\o\sigma$ an anti-holomorphic involution, which in the following
is chosen to be simply  complex conjugation in local
coordinates. It is well known and has been studied in many examples
that this quotient introduces topological defects in the background, so-called
orientifold $O6$-planes, which carry 
tension and charge under the R-R seven-form. 
For dimensional reasons these $O6$-planes wrap 
3-cycles, $\pi_{O6}$, in the Calabi-Yau manifold and since
they preserve one-half of the supersymmetry these 3-cycles
are special Lagrangian (sLag).

To cancel the induced tadpoles of the $O6$-planes one introduces
$D6$-branes, which by themselves wrap in general different
3-cycles. Specifically one introduces $k$ stacks of different branes,
where  on each stack we have $N_a$ $D6$-branes  wrapping the
3-cycle $\pi_a$ and its $\Omega \o\sigma$ image $\pi'_a$.
Then the R-R tadpole cancellation condition takes
the very simple form
\eqn\tadpolea{
\sum_{a=1}^k  N_a\, ( \pi_a + \pi'_a) =
          4\,   \pi_{{\rm O}6} - N_{flux} ,
}
where we have also included the contribution of the flux.
If we also require ${\cal N}=1$ supersymmetry in four dimensions
the branes have to wrap sLag 3-cycles, preserving all the same
supersymmetry.

Similarly to the Type IIB orientifolds discussed in \jan, the homology
group $H_3(M)$ splits into an $\Omega\o\sigma$ even and odd part,
$H_3(M)=H^+_3(M)\oplus H^-_3(M)$. The even part contains real 
3-cycles and the odd part completely imaginary  ones.
Moreover, $\Omega\o\sigma$ exchanges the holomorphic and the
anti-holomorphic 3-forms, so that the volume form
\eqn\volume{ {\rm vol}(M)={i\over 8} \Omega_3\wedge \o\Omega_3 }
is anti-invariant, i.e.  $\Omega\o\sigma:{\rm vol}(M)\to  -{\rm vol}(M)$.
Therefore, the only non-vanishing intersections are between 
3-cycles from  $H^+_3(M)$ and $H^-_3(M)$.

Let us introduce a symplectic basis $(\alpha_I,\beta_I)$ of $H_3(M,\ZZ)$ where
$\alpha_I\in H^+_3(M)$ and $\beta_I\in H^-_3(M)$, so 
that $\alpha_I \cap \beta_J=\delta_{IJ}$ with the other intersection numbers
vanishing. 
We expand the 3-cycles of the branes and the orientifold planes
as
\eqn\basis{\eqalign{
           \pi_a &= \sum_{I=1}^{b_3/2} ( X_{a,I}\, \alpha_I+Y_{a,I}\, \beta_I), \cr 
           \pi'_a &= \sum_{I=1}^{b_3/2} ( X_{a,I}\, \alpha_I-Y_{a,I}\, \beta_I), \cr 
            \pi_{{\rm O}6} &= {1\over 2} \sum_{I=1}^{b_3/2}  L_I\, \alpha_I.\cr
}}
Therefore we get $b_3/2=1+h_{21}$ tadpole cancellation conditions
\eqn\tadpoleb{
       \sum_{a=1}^k  N_a\,  X_{a,I} = L_I - L_{I,{flux}}.
}
Note that in F-theory compactifications the contribution to the
4-form tadpole is given by $L_0=\chi/24$
which can be larger than 32 and of order $10^2$. Even though
in this paper we have a concrete model in mind, we often perform
the analysis for more general values of $L_I$, as we expect
the statistics not to depend too strongly on the details of the model.
The sLag condition 
\eqn\slag{     \Im (\Omega_3)|_{\pi_a}=0  }
takes the form
\eqn\slagb{   \sum_{I=1}^{b_3/2}  Y_{a,I}\, F_I(U)=0.  }
Here $F_I=\int_{\beta_I} \Omega_3$. 
We also have to exclude anti-branes,
leading to the additional condition
\eqn\posa{     \Re (\Omega_3)|_{\pi_a}>0,  }
which can be expanded as
\eqn\posb{   \sum_{I=1}^{b_3/2}  X_{a,I}\, U_I>0. }
The homogeneous complex structure coordinates are defined
as $U_I=\int_{\alpha_I} \Omega_3$. 

In general, two D-branes wrapping some 3-cycles in $M$ have
a non-trivial intersection number given by
\eqn\ints{   I_{ab}=\sum_I X_{a,I}\, Y_{b,I} - Y_{a,I}\, X_{b,I} .}
It is well known that on the intersection locus we 
find $I_{ab}$ chiral multiplets in the bifundamental
representation of the $U(N_a)\times U(N_b)$ gauge symmetry
supported on the pairs of branes. Chiral fermions from the 
$I_{a'a}$ intersections  transform in the symmetric and
antisymmetric representation respectively.

The aim is to count solutions to the tadpole cancellation conditions
\tadpoleb\ under the constraints \slagb\ and \posb.
Here we assume that the contribution from the orientifold planes
and possible other sources are fixed.  
Note that for a specific class of special Lagrangians the wrapping
numbers $X_I$ and $Y_I$ are no longer independent so that
it might indeed happen that the number of solutions
to these equations is finite. In fact, the question about finiteness
of solutions splits into two parts. First, are there finitely many solutions
for fixed $L_I$ and complex structures $U_I$? Second, up to moduli space
identifications of solutions, 
are there only finitely many non-trivial complex structures
allowed for fixed $L_I$? 

Of course we are not only interested in the actual number of solutions
and how they scale with the $L_I$, but also in computing
statistical distributions of physically more interesting quantities like
the rank of the gauge group, the number of families, the number
of Standard-like  models, the number of GUT like models, etc.
It would also be particularly interesting to observe statistical 
correlations between different observables such as
the rank of the gauge symmetry and the number of families, as after
all these might turn out to be the only string theoretic predictions
possible to detect on the string theory landscape.

We will develop techniques to analytically respectively
semi-analytically get a hand on the statistics of these brane solutions.
To be concrete we will discuss our ideas elaborating on the
examples of intersecting D-brane models on the
$T^2$, $T^4\over \ZZ_2$ and  $T^6\over \ZZ_2\times \ZZ_2$ backgrounds.

\subsec{Background on concrete toroidal models}

Let us summarise how these concrete examples as
discussed in the literature fit into the
general scheme introduced in the previous section.
\vskip 0.4cm
\noindent
$\bullet$ $M=T^2$
\vskip 0.2cm
For compactification on $T^2$, a special Lagrangian submanifold is specified
by two wrapping numbers $(n_a,m_a)$ around the fundamental 1-cycles. In this
case these numbers are precisely identical to the numbers $(X_a,Y_a)$ introduced
in the previous section. The supersymmetry condition \slagb\ becomes
$Y_a=0$, which is independent of the complex structure $U=R_2/R_1$
on the rectangular torus and implies that all supersymmetric
branes must lie along the x-axis, i.e. on top of the orientifold
plane.
The second supersymmetry  condition \posb\ becomes $X_a>0$. 
If we did not allow for multiple wrapping (as is usually done
in this framework), there would only exist one supersymmetric brane,
namely the one with $(X,Y)=(1,0)$. Of course this is quite boring and
in order to study the statistics of this model and to develop general
tools we will allow multiple wrappings. Finally, the tadpole
cancellation condition reads
\eqn\tadaa{   \sum_a N_a\, X_a = 16 .} 
\vskip 0.4cm
\noindent
$\bullet$ $M={T^2\times T^2\over \ZZ_2}$
\vskip 0.2cm
In this case, a class of special 
Lagrangian bulk branes is given by so-called factorisable branes,
which are similar to the former case defined by two pairs of
wrapping numbers $(n_{I},m_I)$. The wrapping numbers $(X_I,Y_I)$ with
$I=1,2$ for the $\ZZ_2$ invariant two-dimensional cycles are then
given by
\eqn\wrap{\eqalign{ X_1=n_1\, n_2, \quad\quad X_2=m_1\, m_2, \cr
                    Y_1=n_1\, m_2, \quad\quad Y_2=m_1\, n_2. \cr}}
Note that these branes do not wrap the most general homological class,
for the 2-cycle wrapping numbers satisfy the relation
\eqn\consty{    X_1\, X_2 = Y_1\, Y_2 .}
However, for a more general class we do not  know what the
special Lagrangians are. Via brane recombination it is known
that there exist flat directions in the D-brane moduli space,
corresponding to branes wrapping non-flat special
Lagrangians. To avoid these complications, in this paper
we stick to the well understood branes introduced above. 
However, keep in mind that since there are many more
branes possible, the statistical estimates we are going to
present  in this
paper are actually  a lower bound. We believe and it remains
to be proven that the qualitative features we derive  from
our analysis of this restricted but still large set
of branes are representative. 

The untwisted tadpole cancellation conditions read
\eqn\tadabs{\eqalign{    \sum_a N_a\, X_{a,1} &= 8, \cr
                        \sum_a N_a\, X_{a,2} &= -8.\cr}} 
Note that in contrast to for instance the model discussed
in \rgimpol, for simplicity we are only considering bulk branes without
any twisted sector contribution (see \rbbkl\ for a treatment
of fractional branes in this framework).
Defining $\Omega_2=(dx_1+i U_1 dy_1)(dx_2+i U_2 dy_2)$ the
supersymmetry conditions become
\eqn\susya{\eqalign{   U_1\, Y_1 + U_2\, Y_2 &=0, \cr
                X_1 - U_1\, U_2\, X_2 &>0. \cr}}
The intersection number between two bulk branes has an extra
factor of two
\eqn\intaa{   I_{ab}=2\left( X_{a,1}\, X_{b,2}  +X_{a,2}\, X_{b,1} - Y_{a,1}\, Y_{b,2} -Y_{a,2}\, Y_{b,1}\right) .}
\vskip 0.4cm
\noindent
$\bullet$ $M={T^2\times T^2\times T^2\over \ZZ_2\times \ZZ_2}$
\vskip 0.2cm
Here the situation is very similar.  One introduces
three pairs of wrapping numbers $(n_I,m_I)$, $I=1,2,3$, 
in terms of which the 3-dimensional wrapping numbers can be expressed
as
\eqn\wrapf{\eqalign{ X_0&=n_1\, n_2\, n_3, \quad X_1=n_1\, m_2\, m_3,\quad X_2=m_1\, n_2\, m_3, \quad
    X_3=m_1\, m_2\, n_3, \cr
   Y_0&=m_1\, m_2\, m_3, \quad Y_1=m_1\, n_2\, n_3, \quad Y_2=n_1\, m_2\, n_3, \quad
    Y_3=n_1\, n_2\, m_3 .\cr}}
These are again not independent but satisfy a couple of relations
\eqn\constyy{\eqalign{ X_I\, Y_I &= X_J\, Y_J\quad {\rm for\ all}\ I,J, \cr
                       X_I\, X_J&=Y_K\, Y_L, \cr
                       X_L\, (Y_L)^2&=X_I\, X_J\, X_K, \quad\quad\quad  \cr
                       Y_L\, (X_L)^2&=Y_I\, Y_J\, Y_K \quad {\rm for\ all}\ I\ne J\ne K\ne L\ne I, \cr}}
which by themselves are related. 
The tadpole cancellation conditions read
\eqn\tadab{\eqalign{    \sum_a N_a\, X_{a,0} &= 8, \cr
                        \sum_a N_a\, X_{a,I} &= -8\quad  {\rm with}\ I=1,2,3 .\cr}} 
Let us define the complex structures $\widetilde U_I=R_{2,I}/R_{1,I}$.
Using $\Omega_3=\prod_{I=1}^3 (dx_I + i \widetilde U_I dy_I)$
and defining $U_I=\widetilde U_J \widetilde U_K$ with  $I\ne J\ne K\ne I$
one can express the supersymmetry conditions as
\eqn\susybb{\eqalign{ Y_0 -\sum_I Y_I {1\over U_I} &=0, \cr
                      X_0 -\sum_I X_I {U_I} &>0. \cr}}

\newsec{Statistics of an eight dimensional toy model}

As a simple toy model showing not all the complications of the 
Calabi-Yau case, we discuss intersecting branes on $T^2$.
Recall that in this case there is only one  tadpole cancellation condition which reads
\eqn\tadeight{
       \sum_{a=1}^k  N_a\,  X_a = L. 
}
%The supersymmetry conditions are  just $Y_a=0$ and $X_a>0$, so that in fact
%all branes lie along the x-axes of the $T^2$ and the complex structure
%$U$ is a free parameter. Therefore, if we do not
%allow multiple wrapping, there exists only one solution to \tadeight.
%This is of course boring and therefore  we allow also
%multiple wrapping numbers $X_a>1$. 
The task is now to count all unordered solutions to \tadeight\ with arbitrary number
of stacks $k$. This is a number theoretical problem, for which we
have not found any solution in the literature.

\subsec{Counting partitions}

We proceed by first discussing an easier but apparently
related problem, namely to count unordered solutions (i.e. after dividing out permutations) to the
constraint
\eqn\parta{
       \sum_{a=1}^k  N_a  = L. 
}
This is nothing else than to count the number of partitions of $L$.
In particular, we would be happy with a rough estimate 
for large $L$. To make this paper self-contained let
us discuss in some detail how one can arrive at such an
estimate by using the saddle point approximation (SPA), a well known
and often applied  technique
from the mathematical theory of asymptotic approximations 
\refs{\andrews,\wong}.

Let us discuss two slightly different approaches to this problem.
First note that the problem of counting partitions is  equivalent to counting
solutions to the equation
\eqn\partb{
       \sum_{k=1}^\infty  k\,  n_k  = L, 
}
where the ordering issue is solved automatically.
Now one writes
\eqn\partc{\eqalign{   {\cal N}(L)&=\sum_{sol} 1 = 
                             \sum_{all} \delta_{\sum_{k}\,   k  n_k-L ,0} \cr
                     &={1\over 2\pi i} \oint dq {1\over q^{L+1} }
                     \sum_{n_k=0}^\infty  
                        q^{\sum_{k}\,  k  n_k} \cr
                                &={1\over 2\pi i} \oint dq {1\over q^{L+1}} 
                            \prod_k \left( 1\over 1-q^k \right). \cr
}}
The last line simply extracts the order $L$ term in the generating function
for the partitions, which is the inverse of the well known Dedekind $\eta$-function.
A common method to evaluate the asymptotic expansion of such integrals
is the saddle point approximation. 
Since we intend to apply this method also for counting string vacua,
let us discuss it in some more detail.

First we write
\eqn\partd{ {\cal N}(L)={1\over 2\pi i} \oint {dq} \,
            e^{f(q)} }
with 
\eqn\parte{ f(q)=-\sum_{k=1}^\infty  \log (1- q^k) -(L+1)\, \log q. }
To evaluate this integral one assumes that the main contribution
comes from the saddle points $q_0$, which are determined by the
condition  $df/dq|_{q_0}=0$. Here we assume that there exists only
one saddle point. Otherwise one has to sum over all
saddle  points.
Setting $q=\rho \exp(i\varphi)$ we get 
\eqn\partf{ {\cal N}(L)= {1\over 2\pi} \int_{-\pi}^{\pi} {d\varphi} \, q\,
            e^{f(q)}. }
The leading order saddle point approximation is then simply given by
\eqn\partff{ {\cal N}^{(0)}(L)=  e^{f(q_{0})}.  } 
One can compute the next to leading order approximation by Taylor expansion in $\varphi$
\eqn\taylor{   f(\rho_{0},\varphi)=f(q_{0}) +
    {1\over 2} {\partial^2 f\over \partial \varphi^2}|_{q_{0}}\, \varphi^2
    +\ldots }
and inserting it into \partf.
Using that $(\partial^2 f / \partial \varphi^2)_{q_{0}}=- q^2\, 
             (\partial^2 f / \partial q^2)_{q_{0}}$,
we obtain 
\eqn\partg{ {\cal N}^{(2)}(L)=  {1\over {2\pi}}\,  e^{f(q_{0})} 
   \int_{-q_0 \pi}^{q_0 \pi}  dx \,  e^{-{1\over 2} (\partial^2 f / \partial q^2)_{q_{0}} x^2}}
with $x=q_0\varphi$. Thus for a sufficiently large second derivative
one finally obtains
\eqn\parth{ {\cal N}^{(2)}(L)=  {1\over \sqrt{2\pi}}\, { e^{f(q_{0})} \over 
    \sqrt{{\partial^2 f \over  \partial q^2}|_{q_0} }} .  }

For later use we also give the analogous result  for the saddle
point approximation when more than one variable is involved.
Here one wants  to approximate an integral
\eqn\hda{ {\cal N}(\vec L)={1\over 2\pi i} \oint \prod_{I=1}^n {dq_I} \,
            e^{f(\vec q)}, }
where $f$ in general is of the form 
\eqn\parte{ f(\vec q)=g(\vec q) - \sum_I (L_I+1)\, \log q_I. }
The saddle point is determined by the condition $\nabla f(\vec q)|_{\vec q_0}=0$,
which after Taylor expanding $f(\vec q)$ around the saddle point leads
to the second order saddle point approximation of \hda
\eqn\hbb{ {\cal N}^{(2)}(\vec L)=  {1\over  \sqrt{2\pi}^n}\, { e^{f(\vec q_{0})} 
         \over 
    \sqrt{ \det\left[ \left( {\partial^2 f\over \partial q_I\partial q_J}\right) 
        \right]_{q_0}}}.  }

Now, let us first analytically evaluate the leading order contribution
of \partc.
One notices that for large $L$, the integrand
quickly approaches infinity both for $q<1$ and for $q\simeq 1$, so that
one expects a sharp minimum somewhere very close to $q\simeq 1$.
Therefore we write the integrand as exp$( f(q))$ with
\eqn\saddlea{
       f(q)=\log\left(   \prod_k  {1\over 1-q^k} \right) - (L+1)\, \log q.
}
Close to $q\simeq 1$ we can write the first term in \saddlea\ as
\eqn\saddleb{\eqalign{
       \log\left(   \prod_k  {1\over 1-q^k} \right) & = -\sum_k \log(1-q^k) 
               =\sum_{k,m>0} {1 \over m} q^{km} \cr 
             &\simeq {1\over (1-q)} \sum_{m>0} {1\over m^2} = 
               {\pi^2\over 6} {1\over (1-q)}  \cr
}}
so that \saddlea\ becomes
\eqn\saddlec{
       f(q)\simeq {\pi^2\over 6} {1\over (1-q)} - (L+1)\, \log q.
}
The minimum of this function is approximately at $q_{0}\simeq1-\sqrt{\pi^2\over 6L}$,
so that $f(q_{0})\simeq \pi \sqrt{2L/3}$.
To summarise, a first estimate of the growth of the partitions for large $L$
is given by 
\eqn\saddled{
       {\cal N}(L) \simeq e^{\pi \sqrt{2L/3}}.
}
This is precisely the leading term in the celebrated Hardy-Ramanujan formula
for the asymptotic growth of the number of partitions
\eqn\hardy{
       {\cal N}(L)_{HR} \simeq {1\over 4\, L\, \sqrt 3}\, e^{\pi \sqrt{2L/3}}.
}
As in the more complicated examples we are going to
discuss in this paper an analytic approach is not always feasible,
let us also numerically evaluate the saddle point conditions
and compare the exact, the leading order and the next to leading
order approximations. The resulting three curves for Ln$( {\cal N}(L))$
are shown in Figure 1.
%begin figure
\fig{Number of partitions: The solid line shows the exact
number of partitions, the upper dotted line the leading
order saddle point approximation and the lower
dotted line the second order.}{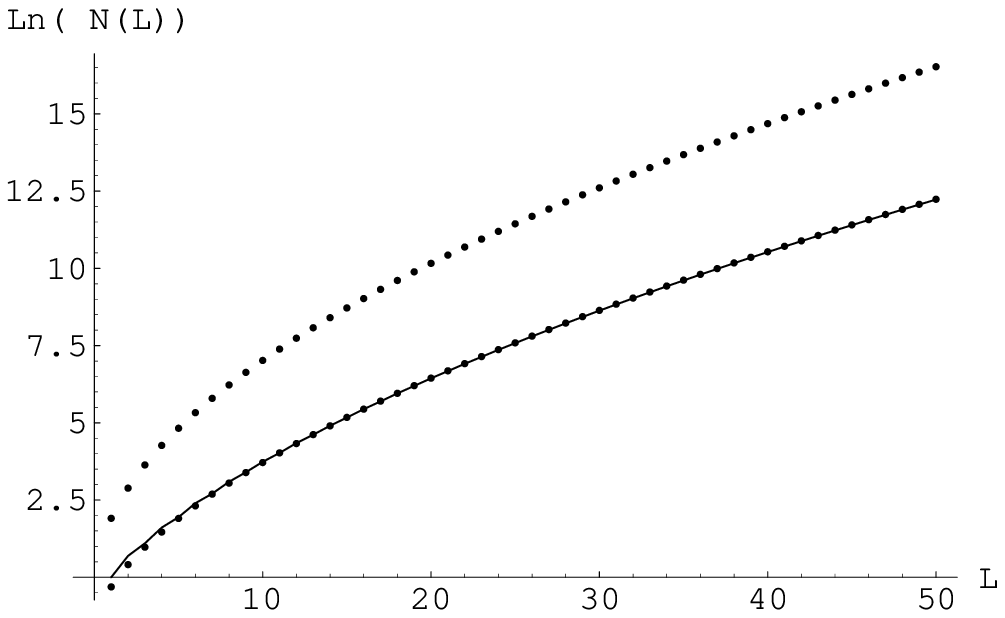}{13truecm}
%end figure
This  shows that the second order saddle point approximation provides
a very good estimate for the behaviour of the number
of partitions. To illustrate this even more we have listed
in Table 1 the values for the number of partitions resulting from 
an exact computation, the Hardy-Ramanujan formula and the second order
saddle point approximation. It is quite impressive  that the numerically
evaluated saddle point method is even better than the Hardy-Ramanujan
formula.

\vskip 0.8cm
\vbox{ \centerline{\vbox{ \hbox{\vbox{\offinterlineskip
\def\tablespace{height2pt&\omit&&\omit&&\omit&&
 \omit&\cr}
\def\tablerule{\tablespace\noalign{\hrule}\tablespace}

\hrule\halign{&\vrule#&\strut\hskip0.2cm\hfill #\hfill\hskip0.2cm\cr
& L &&  ${\cal N}(L)$ &&  ${\cal N}(L)_{HR}$ &&  ${\cal N}^{(2)}(L)_{SPM}$  &\cr
\tablerule
& $1$ &&  $1$ &&  $1.88$ && $0.73$ &\cr
\tablespace
& $2$ &&  $2$ &&  $2.72$ && $1.51$ &\cr
\tablespace
& $3$ &&  $3$ &&  $4.09$ && $2.65$ &\cr
\tablespace
& $4$ &&  $5$ &&  $6.10$ && $4.33$ &\cr
\tablespace
& $5$ &&  $7$ &&  $8.94$ && $6.72$ &\cr
\tablespace
& $6$ &&  $11$ &&  $12.88$ && $10.09$ &\cr
\tablespace
& $7$ &&  $15$ &&  $18.27$ && $14.75$ &\cr
\tablespace
& $8$ &&  $22$ &&  $25.54$ && $21.10$ &\cr
\tablespace
& $9$ &&  $30$ &&  $35.25$ && $29.65$ &\cr
\tablespace
& $10$ &&  $42$ &&  $48.10$ && $41.06$ &\cr
\tablespace
\tablespace
}\hrule}}}} 
\centerline{ \hbox{{\bf
Table 1:}{\it ~~ Comparison of asymptotic approximations for ${\cal N}(L)$ }}} } 
\vskip 0.5cm
\noindent

Recall that in order to get  this nice result we reformulated the problem
of finding solutions to \parta. In view of our original goal
to find unordered solutions to the tadpole cancellation condition \tadeight, it
would be nice  if we could also count the number of partitions using
the original equation
\eqn\partd{
       \sum_{a=1}^k  N_a  = L. 
}
Here the number of stacks is a free parameter over which we have to sum
eventually. Since we are only interested in unordered solutions,
a rough estimate would be that we divide the number of ordered solutions
by $k!$. Of course this too heavily suppresses solutions when
some of the $N_a$ are equal. However, introducing the precise combinatorial
factors is partition dependent and does not simplify the problem.
Therefore, let us see how far we  can get  with the naive $k!$ factor. 
Proceeding now analogously to the former case, we write the total number of
solutions as
\eqn\parte{\eqalign{   \widetilde{\cal N}(L) 
                     &\simeq{1\over 2\pi i} \oint dq {1\over q^{L+1} }
              \sum_{k=1}^\infty {1\over k!}\, \sum_{N_1=1}^\infty \ldots
               \sum_{N_k=1}^\infty  q^{\sum_a \, N_a} =
                     {1\over 2\pi i} \oint dq {1\over q^{L+1} }
                     \sum_{k=1}^\infty  {1\over k!}\, 
                     \left( \sum_{N=1}^\infty  q^{N}\right)^k \cr
                     &={1\over 2\pi i} \oint dq {1\over q^{L+1} }
                     \sum_{k=1}^\infty {1\over k!}\, \left( {q\over 1-q }\right)^k =
                     {1\over 2\pi i} \oint dq {1\over q^{L+1} } 
               \exp\left( {q\over 1-q }\right). \cr
}}
Applying again the saddle point method yields
\eqn\saddlee{
       \tilde f(q)= {q\over 1-q } - (L+1)\, \log q
}
and leads to the final analytic result for the leading order
approximation 
\eqn\saddlef{
       \widetilde{\cal N}(L) \simeq e^{2 \sqrt{L}}.
}
We conclude that in this approximation we get the right growth $e^{\sqrt L}$
and only the numerical coefficient is slightly smaller
\eqn\numer{   {\log {\cal N} \over \log \widetilde{\cal N}}={\pi\over \sqrt 6}
                  \simeq 1.28 .}  
Now we can also numerically determine  the saddle points and compute
the leading and next to leading order results.
These are shown in Figure 2.
%begin figure
\fig{Saddle point approximation: The solid line denotes the exact number
of partitions, the dotted line the leading order SPA
and the stared line the second order. Moreover, we have
shown with triangles the second order approximation with
the analytic factor $1.28$ included.}{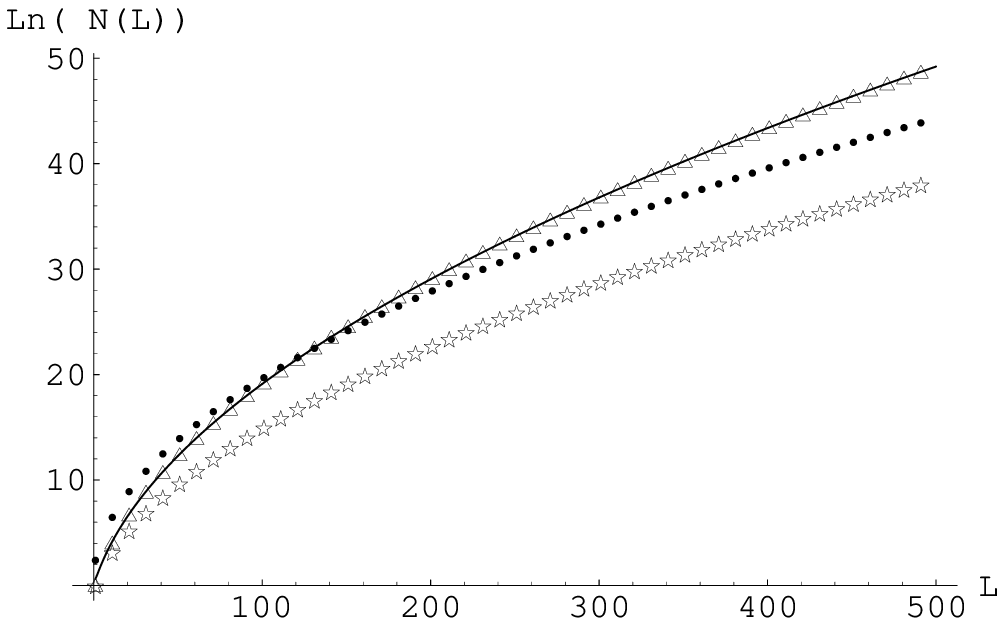}{13truecm}
%end figur
As expected we realize  that for  the second order approximation the curve
 lies below the exact result,
but at least qualitatively gives a reasonable result.
The difference is due to the fact that the $1/k!$ combinatorial factor
is responsible for too large a suppression of partitions with identical
terms. However, note that it was precisely this crude 
factor which allowed us to proceed with the computation in the first place
and to simplify the integrals considerably. 
Keeping in mind that we cannot expect  to obtain always  quantitatively 
completely accurate
results, from now on we apply the saddle point method  for counting
solutions to the tadpole conditions.
What we are mainly aiming at is to get  
qualitatively  correct pictures of the distributions of various
physical quantities. Comparing our results with  a brute force computer search
reveals  that also the quantitative agreement in certain cases
is fairly good. 

\subsec{A comment on counting rigid branes}

The string models we are going to consider in this paper  mostly have
the feature that the D-branes allow for continuous displacements
and Wilson lines, i.e. the cycles they wrap are not rigid. 
For more realistic models it is however desirable to get rid of the
extra adjoint scalars and consider rigid cycles instead 
\foot{We would like to thank M.R. Douglas for
encouraging us during the "String Vacuum Workshop" in Munich  
to study the effects discussed in this section and in section
3.7.}.. 
Then the question arises  whether this drastically changes the statistics 
of the D-brane sector. As an indication that one does not expect this to
happen, at least qualitatively, let us consider the analogous question for
the related problem of counting partitions. 

For rigid branes, one is not allowed to have two stacks with identical wrapping
numbers. Analogously, we would like to count unordered partitions where each
term can only appear once. This is like Fermi statistics and the
exact expression for this number ${\cal Q}(L)$ takes the familiar form
\eqn\ferma{   {\cal Q}(L)={1\over 2\pi i} \oint dq {1\over q^{L+1}} 
                            \prod_k \left( 1+q^k \right).
}
Let us  write the integrand as exp$( f(q))$ with
\eqn\fermb{
       f(q)=\log\left(   \prod_k  (1+q^k) \right) - (L+1)\, \log q.
}
Close to $q\simeq 1$ we can write the first term in \fermb\ as
\eqn\fermc{\eqalign{
       \log\left(   \prod_k  (1+q^k) \right) & = \sum_k \log (1+q^k) 
               =\sum_{k,m>0} {(-1)^{m+1} \over m} q^{km} \cr 
             &\simeq {1\over (1-q)} \sum_{m>0} {(-1)^{m+1}\over m^2} = 
               {\pi^2\over 12} {1\over (1-q)} . \cr
}}
Proceeding in the same way as above, we find that the minimum of this function is 
approximately at $q_{0}\simeq1-\sqrt{\pi^2\over 12L}$, so that
\eqn\fermd{
       {\cal Q}(L) \simeq e^{\pi \sqrt{L/3}}.
}
Therefore, in the leading order saddle point approximation ${\cal N}(L)$
and  ${\cal Q}(L)$ show qualitatively the same exponential growth with 
$\sqrt{L}$ and 
\eqn\ferme{   {\log {\cal N} \over \log {\cal Q}}={\sqrt 2}.}
The leading order saddle point approximation gives the exponential term
in the well known asymptotic expansion for  ${\cal Q}(L)$
\eqn\fermf{
       {\cal Q}(L) \simeq {1\over 4\cdot 3^{1\over 4}\, L^{3\over 4}}\, e^{\pi \sqrt{L/3}}.
}
We consider these results as evidence that  qualitatively the
distributions of physical observables do  not change for rigid branes.

\subsec{Counting tadpole solutions}

Now that we have convinced ourselves that, from a pragmatic point of
view, the second approach from section (3.1) gives already a good estimate,
we want to apply the same method to count the solutions of 
the tadpole condition
\eqn\tadea{
       \sum_{a=1}^k  N_a\,  X_a = L. 
}
By the same reasoning we compute 
\eqn\tadeb{\eqalign{   {\cal N}(L) 
                     &\simeq{1\over 2\pi i} \oint dq {1\over q^{L+1} }
              \sum_{k=1}^\infty {1\over k!}\, 
                \sum_{N_1=1}^\infty\sum_{X_1=1}^L \ldots
                  \sum_{N_k=1}^\infty\sum_{X_k=1}^L 
                           q^{\sum_a  N_a X_a} \cr
           &={1\over 2\pi i} \oint dq {1\over q^{L+1} }
                     \sum_{k=1}^\infty {1\over k!}\, 
                  \left(\sum_{X=1}^L {q^X\over 1-q^X} \right)^k \cr
}}
so that the saddle point function $f$ reads
\eqn\tadec{
       f(q)= \sum_{X=1}^L {q^X\over 1-q^X } - (L+1)\, \log q. 
}
Close to $q\simeq 1$ we find the analytical expression
\eqn\taded{\eqalign{  f(q)&\simeq {1\over 1-q}\sum_{X=1}^L {1\over X} -L\, 
                \log q   \simeq {\log L\over 1-q} -L\, \log q .\cr 
}}
In this approximation the saddle point is at $q_{min}=1-\sqrt{\log L\over L}$, so that
\eqn\tadee{   {\cal N}(L)\simeq e^{2\sqrt{ L\log L}} .} 
A rough intuitive understanding  of this result can be gained  as follows.
In order to solve \tadea\ one first divides $L$ into its partitions and
then one writes each term as a product of two positive integer numbers.
We know already that the number of partitions scales like $e^{2\sqrt{ L}}$.
On the other hand it is known in number theory that the function $\sigma_0(n)$ 
of divisors of an integer $n$ has the property
\eqn\divi{   {1\over L} \sum_{n=1}^L \sigma_0(n)\simeq \log L + (2\gamma_E -1), }
where $\gamma_E$ denotes the Euler-Mascheroni constant. We think that this explains
the appearance of the $(\log L)$ factor in \tadee. Of course, further L-dependent 
corrections will arise from the sub-leading second  order terms in the SPA.  

In order to test these results  one can now compare the three different
approaches  to  unravel the statistics of solutions of
the tadpole equation.  

\item{$\bullet$} First the {\it numerical approach}, where  one performs for varying $L$ a 
brute force 
computer search for all solutions
to the tadpole conditions. This is the most time consuming and
less elegant method but it really  gives  the exact number
of solutions. 

\item{$\bullet$} Second, we can use the analytical  saddle point function $f(q)$ 
\tadec\ and using MATHEMATICA determine numerically  
the saddle points for varying $L$. Here only the sum over all possible 
wrapping numbers for just a single stack of branes occurs, 
which a computer can easily handle in O(1) seconds.
We call this approach  {\it semi-analytic} and it will be the most
powerful one for the more complicated 6D and 4D models.

\item{$\bullet$} Third, one can try to estimate the
locations of the saddle points and derive an analytic expression
for ${\cal N}(L)$. In the easy example above it was possible 
to do so, but in more complicated examples it gets more
and more complicated. We call this approach the {\it analytic}
one.

In Figure 3 we display the result for the total number of solutions
to the tadpole cancellation condition for varying number of $L$.
Here we compare the result of the exact numerical analysis with
the outcome of the second order semi-analytic SPA.
\vskip 0.2cm
%begin figure
\fig{Number of solutions for varying $L$. The dotted line denotes the result
of the exact analytic computer search and the solid line the
outcome of the second order SPA.}{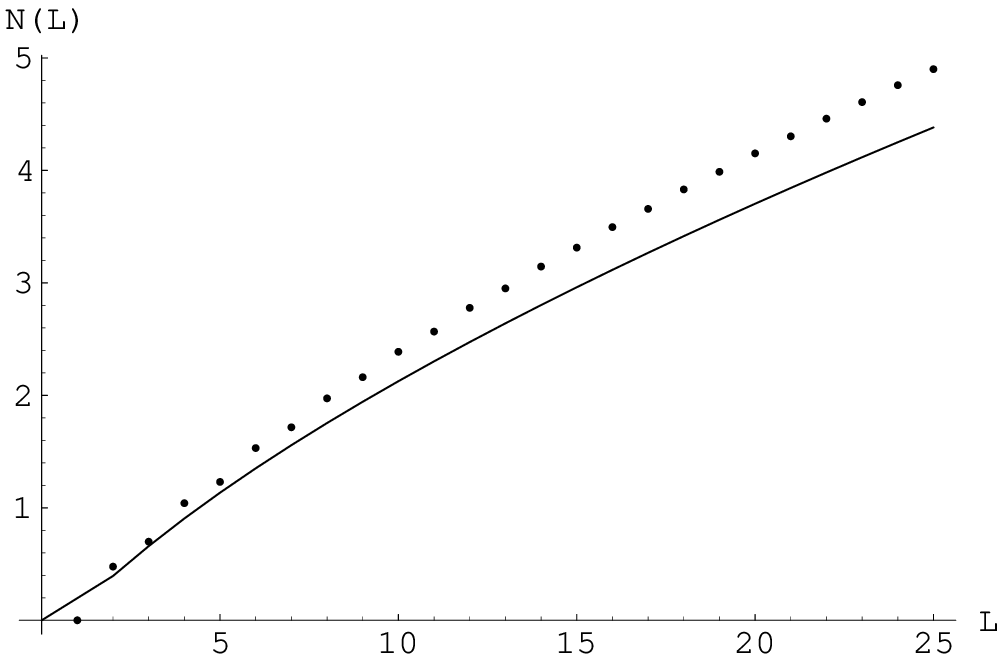}{10truecm}
%end figure
As expected the saddle point curve lies below the exact results, but
nevertheless captures all of the qualitative features.

\subsec{Probability of $SU(M)$ gauge symmetry}

We are also interested in the percentage of models containing
at least one $SU(M)$ gauge factor. With the same methods as above this
can be written as
\eqn\proba{\eqalign{   P(M) 
                     &\simeq {1\over 2\pi i\, {\cal N(L)} } \oint dq\,
               {1\over q^{L+1} }
               \sum_{k=1}^\infty {1\over (k-1)!}\, 
                  \left(\sum_{X=1}^L {q^X\over 1-q^X} \right)^{k-1}\,
                  \sum_{X=1}^L \sum_{N=1}^\infty q^{NX} \delta_{N,M} \cr
               &= {1\over 2\pi i\, {\cal N(L)} } \oint dq\,
               {1\over q^{L+1} }\,
               \exp\left( \sum_{X=1}^L  {q^X\over 1-q^X} \right)\, q^M 
                   \left({1-q^{ML}\over 1- q^M}\right).\cr
}}
The corresponding saddle point function reads
\eqn\probb{
       f(q)= \sum_{X=1}^L {q^X\over 1-q^X } + \log\left[q^M 
                   \left({1-q^{ML}\over 1- q^M}\right) \right]- (L+1)\, \log q. 
}
We can either numerically search for saddle points of this function or
we can observe that for $M\ll L$ the second, $M$ dependent term in 
\probb\ is just a small perturbation. In this case we expect
that in leading order the saddle point does not change, so that we have only
to evaluate \probb\ at $q_{min}=1-\sqrt{ \log L\over L}$. 
Doing this for large $L$ one expects that the probability
to find an $SU(M)$ gauge factor scales like
\eqn\probc{
      P(M)\simeq \exp\left( -\sqrt{\log L \over L} M \right)
}
with the normalisation chosen  appropriately. 
In Figure 4 we have shown the distribution of probabilities for at least 
one $SU(M)$ factor for L=25.
%begin figure
\fig{Probability of at least one $SU(M)$ gauge factor for L=25. The dotted line shows
the exact computer search result and the solid line the second order SPA.}{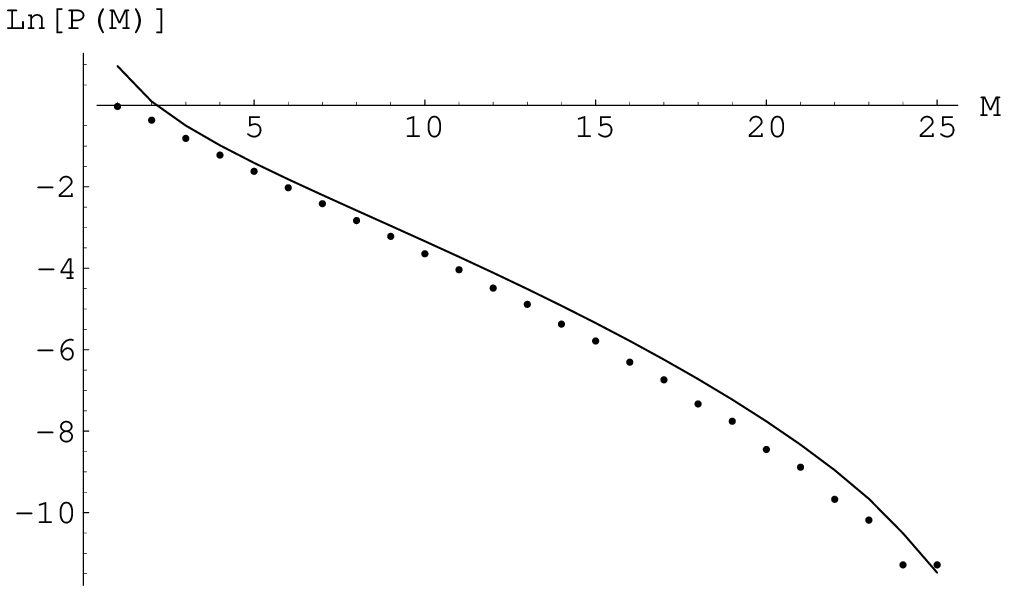}{10truecm}
%end figure
We notice  that the exact and the semi-analytic computation nicely agree.
Since we are computing a normalized quantity the quantitative
agreement is even better than for the total number of states.

Analogously,  the probability to find a product gauge symmetry $\prod_i SU(M_i)$ 
for $\sum_i M_i\ll L$  is
\eqn\probd{
      P(\vec M)\simeq \exp\left( -\sqrt{\log L \over L} \sum_i M_i \right),
}
which shows that in this approximation it only depends on the rank
of the gauge symmetry. Therefore, in this regime
we have mutual independence of the occurrences of $SU(M)$ gauge factors, i.e.
\eqn\indep{ P(\vec M)=\prod_i P(M_i) .}
This behaviour is expected to break down for $\sum_i M_i\simeq L$.

\subsec{The average rank}

Another gauge theoretic quantity we can compute is the
expectation value of the rank of the gauge group
in the ensemble of solutions to the tadpole
cancellation condition.
The integral for the saddle point approximation
can be simplified and written in the form
\eqn\ava{\eqalign{   \la {\rm rank}(G) \ra_{L} 
                  &\simeq{1\over {\cal N}(L)\, 2\pi i} \oint dq {1\over q^{L+1} }
              \sum_{k=1}^\infty {1\over k!}\, 
                \sum_{N_1=1}^\infty\sum_{X_1=1}^L \ldots
                  \sum_{N_k=1}^\infty\sum_{X_k=1}^L 
                           q^{\sum_a  N_a X_a} \left(\sum_a N_a\right)\cr
                    &\simeq {1\over {\cal N}(L)(2\pi i)} 
    \oint   {dq}  \, \exp\Biggl( \sum_{X=1}^L {  q^{X}
                \over 1- q^{X}} +\log\left( \sum_{X} { q^{X}
                \over (1- q^{X})^2 }\right) - (L+1)\log q \Biggr). \cr 
}}
For $L=25$ we obtain the  curves in Figure 5.
%begin figure
\fig{The average rank. 
The dotted line shows the exact computer search result and
the solid  line the second order SPA.}{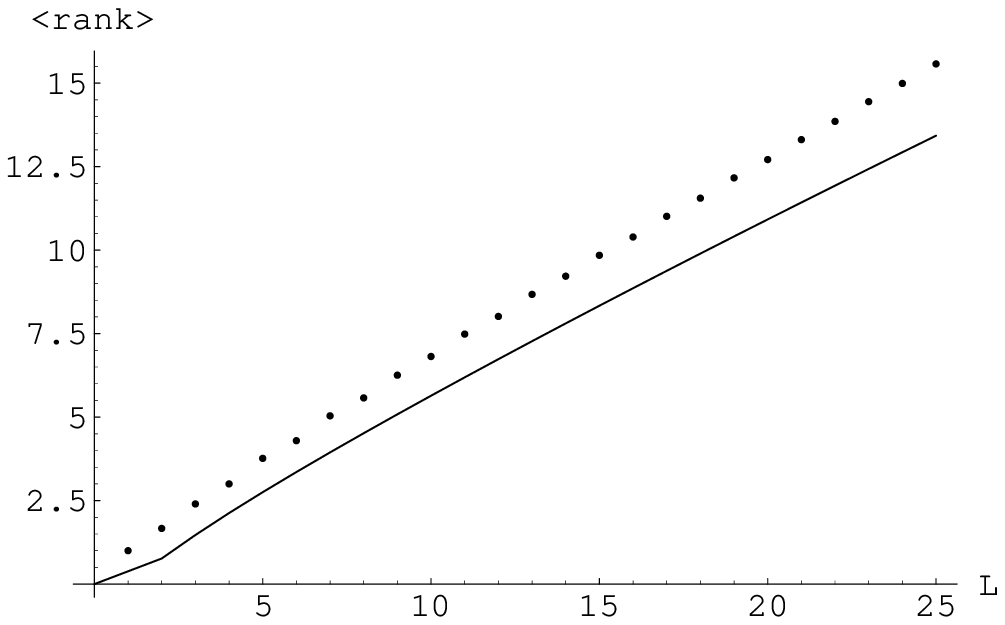}{10truecm}
%end figure
Apparently, the SPA provides a very good estimate
and the average rank scales like $L/2$.

\subsec{The rank distribution}

Next we would like to investigate the probability to get
a gauge group of rank $r$. This means that we also have to implement 
the constraint 
\eqn\ranka{
\sum_{a=1}^\infty N_a =r,
}
which we again do by writing the Kronecker delta as a contour integral. Thus,
\eqn\rankb{\eqalign{   P(r) 
                     &\simeq{1\over 2\pi i\, {\cal N}(L)} 
        \oint dq {1\over q^{L+1} }  \oint dz {1\over z^{r+1} }
              \sum_{k=1}^\infty {1\over k!}\, 
                \sum_{N_1=1}^\infty\sum_{X_1=1}^L \ldots
                  \sum_{N_k=1}^\infty\sum_{X_k=1}^L 
                           q^{\sum_a  N_a X_a}\, z^{\sum_a  N_a} \cr
           &={1\over 2\pi i{\cal N}(L)} \oint dq {1\over q^{L+1} }
           \oint dz {1\over z^{r+1} }
             \exp\left( \sum_{X=1}^L  {z\, q^X\over 1-z\, q^X} \right)\cr
}}
with the corresponding saddle point function
\eqn\rankc{
       f(q,z)= \sum_{X=1}^L {z\, q^X\over 1-z\, q^X } - (L+1)\, \log q-(r+1)\, \log z. 
}
Numerically determining the saddle point in the two variables $q$ and
$z$ we find the Gaussian like distribution shown in Figure 6, where we have
also displayed the exact numerical result. Again the qualitative
agreement is good, the exact curve simply reflects
the number theory behind this distribution.
%begin figure
\fig{The rank distribution for $L=25$. The dots show
the exact results and the solid line the distribution 
coming from the SPA.}{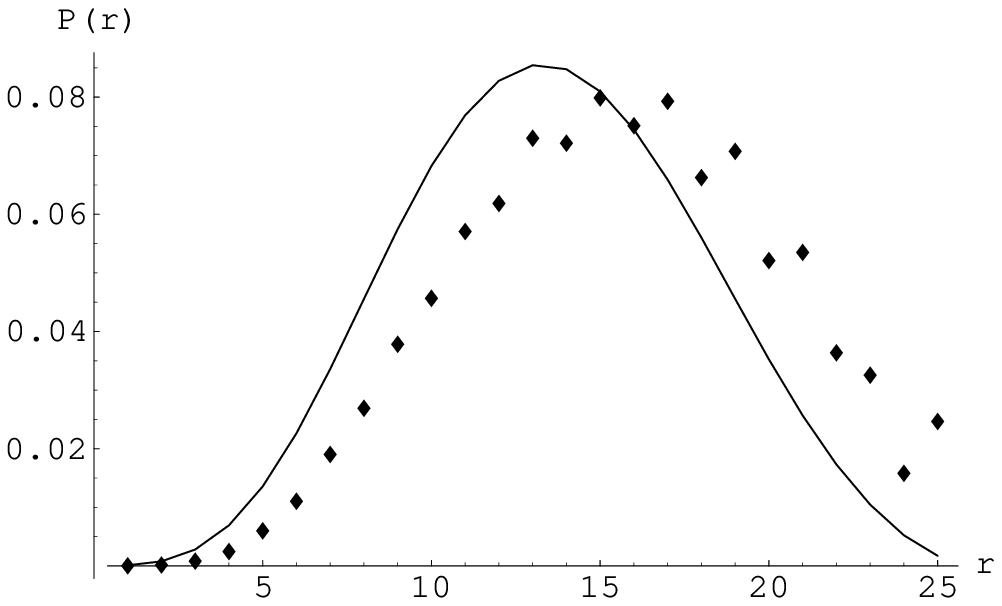}{10truecm}
%end figure
Let us give a heuristic argument for the appearance of the
Gau\ss\ curve. We have already found that the likelihood
for the occurrence of a gauge group of rank $r$ is
$\exp(-\sqrt{\gamma^{-1}} r)$ with $\gamma=L/\log L$. 
In addition, we have the partition of $r$ ways to distribute 
the $N_a$. Therefore, the probability to find rank $r$ 
should behave like
\eqn\rankd{
   P(r)\simeq  \exp\left( 2\sqrt r - \gamma^{-1/2} r\right)\simeq 
              \exp\left( -  {(\sqrt{r}-\sqrt\gamma)^2 \over \sqrt\gamma } \right) .
}
From this argument we expect the maximum of the distribution
to be at $r_{max}=L/\log L$. The width of the Gaussian is
$\Delta r=\sqrt{L/\log L}$, so that the relative width
$\Delta r/r_{min}=\sqrt{\log L/L}$ decreases for larger
$L$.

Since supersymmetry requires $Y_a=0$ for all D-branes, in this
case the intersection number $I_{ab}$ between pairs
of branes always vanishes. In this respect our simple
toy model is not general enough, so that now we move forward
and generalise the methods developed in this section
to the more interesting 6D and 4D models.

\subsec{A comment on counting quantum vacua}

In this paper we mainly consider the statistics of classical vacua
of intersecting D-brane models. In \DouglasUM\ it was suggested that one
should also count quantum vacua, in the sense that the number
of such  vacua of a four-dimensional  ${\cal N}=1$ supersymmetric gauge theory is
given by the Witten index \WittenDF. For a pure $SU(N)$ gauge theory
this index is given by $N$. In this section we would like to briefly discuss
what happens  when we take such an $N$-fold degenerary into account
for our toy model. 

The number of quantum vacua would be given by
\eqn\quana{\eqalign{   {\cal N}_Q (L) 
                     &\simeq{1\over 2\pi i} \oint dq {1\over q^{L+1} }
              \sum_{k=1}^\infty {1\over k!}\, 
                \sum_{N_1=1}^\infty\sum_{X_1=1}^L \ldots
                  \sum_{N_k=1}^\infty\sum_{X_k=1}^L 
                           \left( \prod_a N_a\right)\,  q^{\sum_a  N_a X_a} \cr
           &={1\over 2\pi i} \oint dq {1\over q^{L+1} }
                     \sum_{k=1}^\infty {1\over k!}\, 
                  \left(\sum_{X=1}^L {q^X\over (1-q^X)^2} \right)^k  \cr
}}
so that the saddle point function $f$ reads
\eqn\quanb{
       f(q)= \sum_{X=1}^L {q^X\over ( 1-q^X)^2 } - (L+1)\, \log q. 
}
Close to $q\simeq 1$ we find the analytical expression
\eqn\quanc{\eqalign{  f(q)&\simeq {1\over (1-q)^2}\sum_{X=1}^L {1\over X^2} -L\, 
                \log q   \simeq {\pi^2 \over 6}\, {1\over  (1-q)^2} -L\, \log q .\cr 
}}
In this approximation the saddle point  is at 
$q_{min}=1-\left({\pi^2\over 3\, L}\right)^{1/3}$ so that
the leading order saddle point approximation reads
\eqn\quand{   {\cal N}_Q (L)\simeq \exp \left( \alpha\,  L^{2\over 3} \right) }
with $\alpha={\pi^2+3\over 3^{1/3}\cdot \pi^{4/3}}\simeq 1.94$. 

In Figure 7 we show the numerical leading order saddle point approximation 
confirming the
scaling \quand. This admittedly   very simplified picture shows that
taking the quantum degeneracy into account induces a slight change in the
exponential growth 
\eqn\quand{   {\log {\cal N}_Q \over  \log {\cal N}}\simeq  L^{1\over 6} .}
%begin figure
\fig{The total number of quantum vacua for  $L=50$. The dots show
the leading order saddle point approximation and the solid line the curve
$2.2\cdot L^{2\over 3}/\log 10$.}{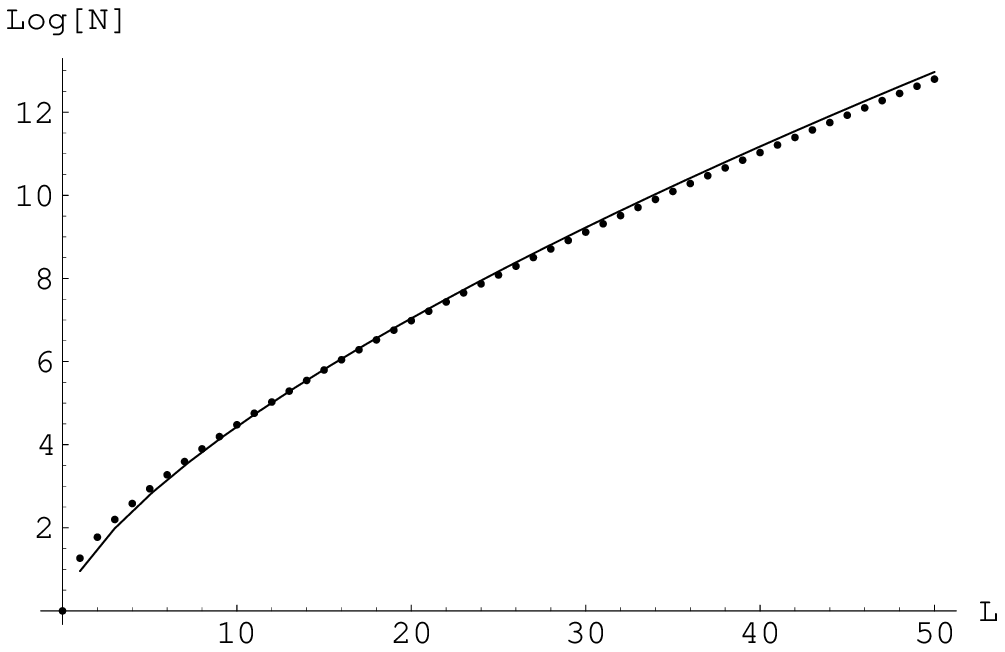}{10truecm}
%end figure

\newsec{Statistics of a 6D model}

In this section,  employing  analogous methods as developed
in the previous one we investigate the statistics of the 
six-dimensional intersecting brane model on the
orientifold  $T^4/\ZZ_2$.
The two main new features as compared to the 8D case are the complex structure dependence of the 
supersymmetry conditions and the possibility of chiral supersymmetric
models. To begin with, let us show that the model
actually admits only  a finite number of inequivalent  solutions to the
tadpole cancellation conditions.

\subsec{Finiteness of solutions}

Recall the general features of this model mentioned in section 2.
Besides these we will show in appendix A that the relative coprime
property of the wrapping numbers $(n_I,m_I)$ translates into the following
relation for the wrapping numbers $(X_I,Y_I)$ 
\eqn\gcda{\eqalign{ {\rm gcd}(X_{1}, Y_{2}) \, {\rm gcd}( X_{2}, Y_{2}) =  Y_{2}.}}
Here we have assumed  
non-vanishing $X_{I}$ and $Y_{I}$. In configurations with some wrapping numbers vanishing
we only admit $(X_1,X_2)\,\in\,\{(0,-1),(1,0)\}$ with $Y_1=Y_2=0$ to exclude multiple wrapping.

In order to decide if the number of stable solutions is finite, we need to
extract the  restrictions on 
possible choices of $U_I$ imposed by supersymmetry \susya\ and 
the tadpole cancellation conditions \tadabs.
It is clear that the supersymmetry conditions can only have 
non-trivial solutions if $U_1/U_2\in \IQ$.
Now, writing $U_1/U_2 = u_1/u_2$ for coprime integers $u_1, u_2$, we find from \susya\ that 
$Y_{1} = u_2 \, y_{1}$ for some integer $ y_{1}$, in terms 
of which \consty\ reads
\eqn\relsixa{\eqalign{
 X_{1} \,X_{2} = - u_1 \,u_2 \,(y_{1})^2.
}} 
It follows that at least 
one of the $X_I$ has to be non-vanishing, in which case the second supersymmetry
 condition becomes vacuous 
once  \tadabs\ and \susya\ are satisfied. 
Note furthermore that due to the reflection symmetry of the 
wrapping numbers we will stick in the following to the case that $X_{a,1}> 0$ and 
change $X_{2} \to - X_{2}$ so that all $X_{I}$ are now positive.

To summarise, we have to analyse all possible solutions to the equations
\eqn\tcsixc{\eqalign{\sum_{a=1}^k N_a \, X_{a,1} &= L_1,\cr
         \sum_{a=1}^k N_a \,{X}_{a,2} &= L_2}} 
for non-negative integers 
$(X_{a,1}, {X}_{a,2}) \neq (0,0)$  
such that
\eqn\roota{ 
\a:=\sqrt{ {X_{a,1} \, {X}_{a,2}}\over {u_1 \, u_2}} \in \ZZ
}
and
\eqn\gcdb{
{\rm gcd}(X_{a,1}, u_1 \, \a) \, {\rm gcd}( {X}_{a,2}, u_1 \, \a) = u_1 \, \a.
}
This means that for fixed complex structures and bounds $L_1$ and $L_2$ the
number of supersymmetric solutions to the tadpole cancellation conditions
is finite.
We denote the set of wrapping numbers $X_1, X_2$ satisfying  \roota\ for fixed complex
structure as $S_U$. Note that we allow for different stacks having the same
wrapping numbers, which means that the branes do have  open string moduli
corresponding to the position and continuous Wilson lines of the branes. 

It remains to be seen that also only a finite number of complex structures are allowed.
In this respect we realise that for given $L_1,L_2$  solutions with at least 
one stack of branes where  $X_1>0$ and   
${X}_2 >0$ are possible only for those complex structures satisfying
\eqn\com{\eqalign{ u_1 \, u_2 \leq L_1\,L_2.}}
In those cases, the factorisability of the branes 
and supersymmetry restrict the admissible ratios of complex structures to only a finite number of 
possibilities. 
Of course, we can still rescale $U_1$ and $U_2$ by the same factor, which
implies that in these models only one of the two complex structure moduli is
fixed and one is left as a free parameter. 

From what we said, it is appropriate  to distinguish two classes of models.
In the first class we have $u_1\, u_2 \leq L_1\, L_2$ and  both 
branes satisfying ( $X_1>0$ and $X_2>0$)  and branes satisfying 
($X_1=0$ or $X_2=0$) are present.
If a solution to the tadpole cancellation condition contains
at least one  brane with ( $X_1>0$ and $X_2>0$), then one complex structure modulus
is fixed, namely $U_1/U_2$. If the solution only contains branes of the other
type, then both complex structures are free parameters.
The second class consists of configurations where $u_1\, u_2 > L_1\, L_2$ and only branes
satisfying ($X_1=0$ or $X_2=0$) are present. For  
coprime wrapping numbers these are only two branes. Apparently, all the solutions to
the tadpole cancellation conditions we can get are already contained
in the first class, so that we can dismiss the second class.
The upshot of this discussion is that, if we count solutions to the tadpole 
cancellation conditions modulo   moduli space identifications, then
the overall number of solutions to the tadpole cancellation
conditions is finite.
 
Given the supersymmetry constraints above, one can simplify further
the intersection number
\eqn\intaa{   I_{ab}=-2\left(X_{a,1}\, X_{b,2}  +X_{a,2}\, X_{b,1} + 
            Y_{a,1}\, Y_{b,2} + Y_{a,2}\, Y_{b,1}\right) }
between two 3-cycles. After some little algebra one arrives at the expressions
\eqn\intaa{\eqalign{   I_{ab}&=-2\,\left(\sqrt{X_{a,1}\, X_{b,2} }-\sqrt{X_{a,2}\, X_{b,1}}\right)^2 ,\cr
                       I_{a'b}&=-2\, \left(\sqrt{X_{a,1}\, X_{b,2} }+\sqrt{X_{a,2}\, X_{b,1}}\right)^2,  \cr}}
which immediately imply $I_{aa}=0$ and $I_{a'a}=-8 X_{a,1}\, X_{a,2}$.
Apparently, the intersection number is negative definite. This is
well in accord with the expectation from ${\cal N}=1$ supersymmetry in six dimensions,
namely that all bifundamental matter fields transform in hypermultiplets and not in vectormultiplets.
For the intersection of the orientifold plane with a D-brane we get
\eqn\intab{  I_{Oa}=-4\, (X_{a,1} + X_{a,2}). }

\subsec{Counting tadpole solutions}

Now we proceed completely analogously to the 8D example and first compute
the total number of solutions to the tadpole cancellation conditions 
for fixed complex structures, $U_1, U_2$.
Again in the occurring integrals the sums over the numbers of branes $N_a$ can be
carried out explicitly and yield geometric series. 
One can write the final result as
\eqn\sixa{\eqalign{   {\cal N}(L_1,L_2) 
                     \simeq {1\over (2\pi i)^{2}} 
    \oint   {dq_1} \, {dq_2} \,
                \exp\biggl(&\sum_{X_I\in S_U} {  q_1^{X_1}\, q_2^{X_2}
                \over 1- q_1^{X_1}\, q_2^{X_2}}  \cr
          &-(L_1+1)\log q_1 - (L_2+1)\log q_2\biggr). \cr 
}}
The asymptotic growth of this expression can be deduced by the saddle point
approximation now in two variables $q_1,q_2$.
In Figure 8 we show the result of a numerical evaluation of this
saddle point for the case of $L_2=8$ and varying $L_1$.
As expected the curve for coprime wrapping numbers lies
under the line with multiple wrappings allowed.
%begin figure
\fig{The total number of models for $L_2=8$, varying $L_1$ and $(u_1=1,u_2=1)$. The upper dots show the
distribution with multiple wrapping numbers allowed and the lower
stared line the case with coprime wrapping numbers.}{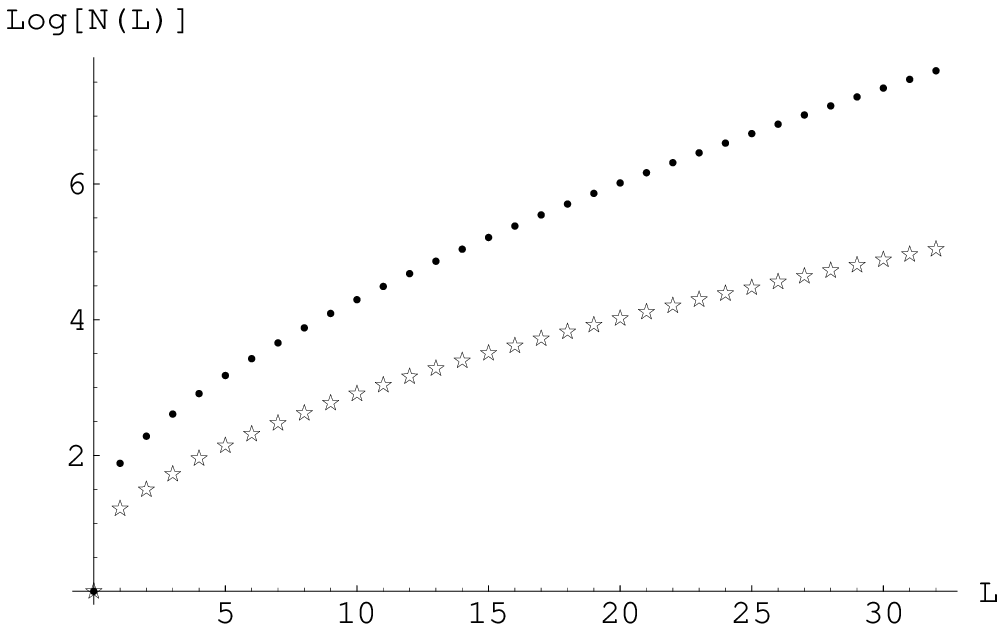}{10truecm}
%end figure
For $L_2=16$ we get the result shown in Figure 9.
Quite surprisingly for fixed $L_2$ this still scales like $\exp(\sqrt{L_1\log L_1})$
so that the  8D toy model gave already a good impression of how
the situation is for more realistic models. 
%begin figure
\fig{The total number of models for $L_2=16$ and $(u_1=1,u_2=1)$}{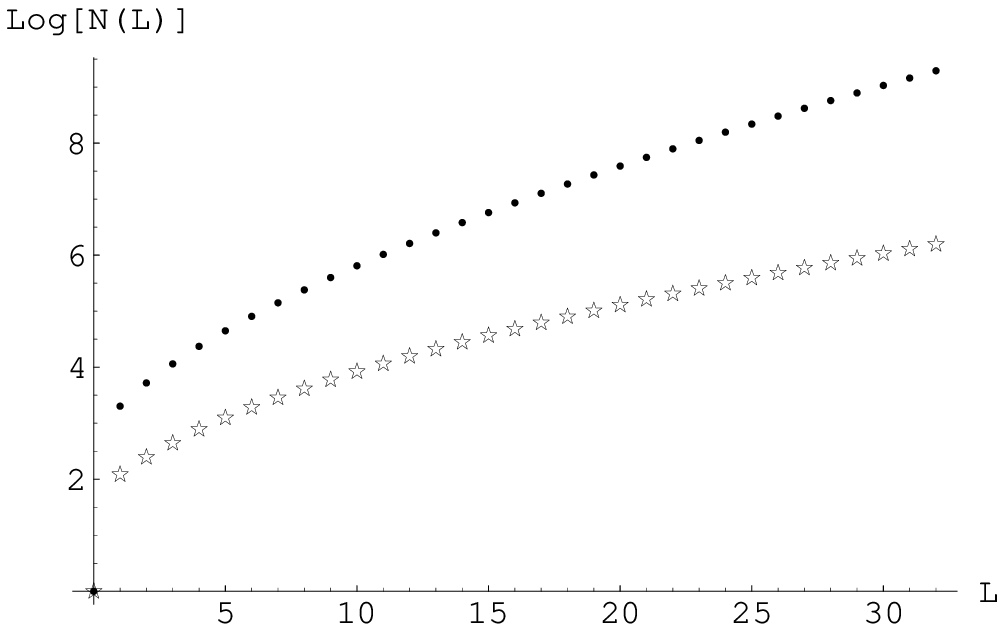}{10truecm}
%end figure

For the case of coprime wrapping numbers we have also carried
out a brute force computer search for all solutions to
the tadpole conditions. The resulting curve for the total number
of solutions is plotted in Figure 10.
As expected from the too strong $1/k!$ suppression, the saddle point
approximation lies below the exact curve. However, qualitatively we 
find the same scaling and if we multiply the saddle point result
by the factor $1.28$ found for the number of partitions in section
3.1, we indeed find quite good agreement.
%begin figure
\fig{The total number of models for $L_2=8$ and $(u_1=1,u_2=1)$.
The dotted line is the result of the exact computer search, the solid line the prediction
from the second order SPA and the dashed
line the latter distribution scaled with the factor $1.28$.}{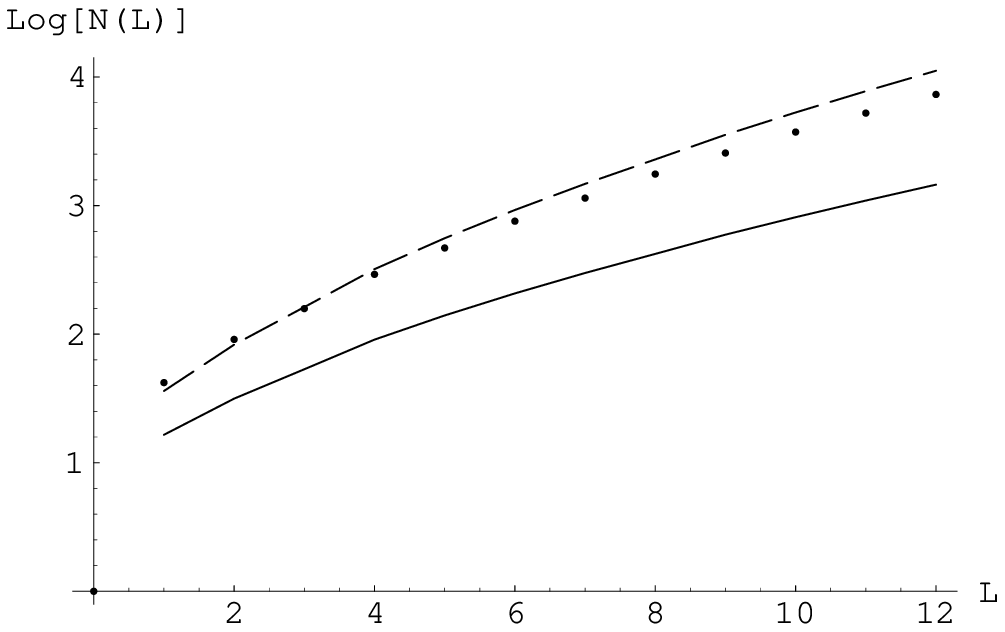}{10truecm}
%end figure

\subsec{Probability of $SU(M)$ gauge symmetry}

Next for fixed $L_I$ we investigate the probability to find at least one $SU(M)$ gauge factor,
which leads to the expression
\eqn\sixa{\eqalign{   P(M,L_I) 
                     \simeq {1\over {\cal N}(L_1,L_2)(2\pi i)^{2}} 
    &\oint   {dq_1} \, {dq_2} \,
                \exp\Biggl(\sum_{X_I\in S_U} {  q_1^{X_1}\, q_2^{X_2}
                \over 1- q_1^{X_1}\, q_2^{X_2}} \cr
           &+\log\left( \sum_{X_I\in S_U} 
              q_1^{M X_1} q_2^{M X_2} \right) - (L_1+1)\log q_1 - (L_2+1)\log q_2\Biggr). \cr 
}}
In Figure 11 we show the resulting distribution for $L_1=8$ and $L_2=8$ and
realise that as in the 8D case it still decreases exponentially with $M$.
%One realizes that as in the 8D case one gets a scaling $\exp(-\kappa M)$, which corresponds in the 
%figure to a straight line. 
%begin figure
\fig{The probability for finding at least one $SU(M)$ gauge factor for $L_1=L_2=8$ and complex 
structure $(u_1=1,u_2=1)$. The dotted line is the result of the exact computer search for
coprime wrapping numbers, which has to be compared with the solid line showing
the second order SPA.}{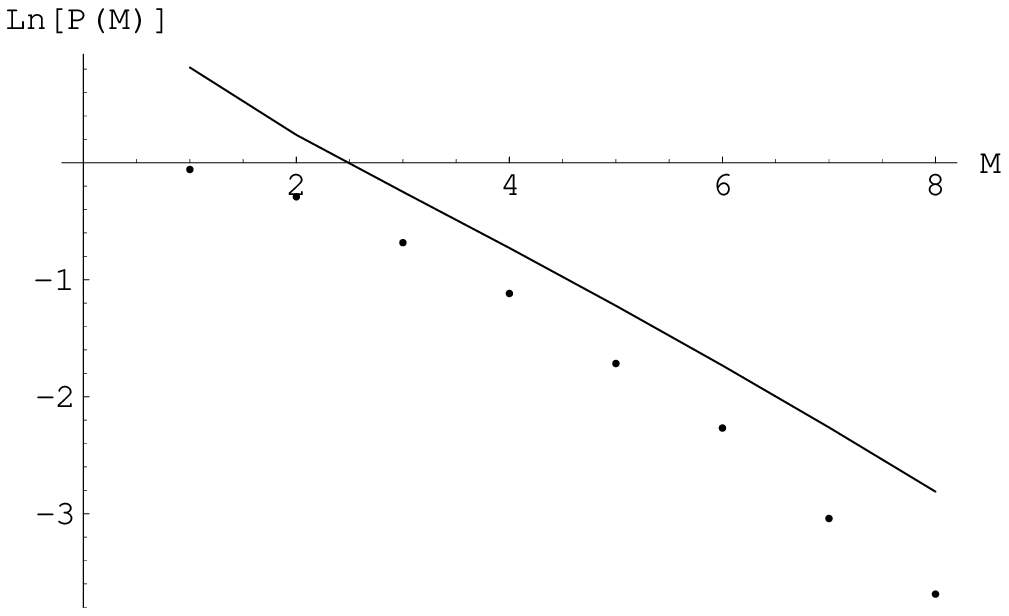}{10truecm}
%end figure
For the choice  $L_1=16$ and $L_2=16$ the distribution looks like 
shown in Figure 12. For $\sum_i M_i\ll L_1+L_2$ we again find that the occurrence of a gauge factor
$\prod_i SU(M_i)$ is given by the product of the probabilities for the occurrence of 
each factor, $P(\vec M)=\prod_i P(M_i)$. 
%begin figure
\fig{The probability for finding at least one $SU(M)$ gauge factor for $L_1=L_2=16$. The dotted
line is the SPA for non-coprime wrapping, the stared line gives the result for
coprime wrapping numbers.}{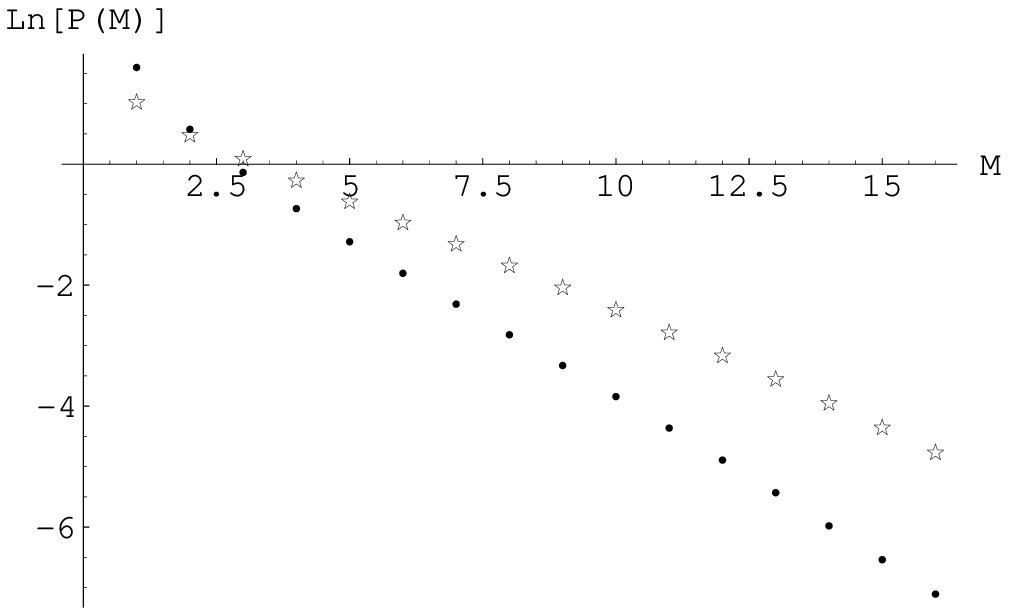}{10truecm}
%end figure

\subsec{The average rank}

In this section we would like to compare the expectation value
of the  rank of the gauge group 
for fixed
$L_2$ and varying $L_1$ as computed from the exact computer search and the
second order SPA.
The integral for the latter
can be simplified and written in the form
\eqn\ava{\eqalign{   \la {\rm rank}(G) \ra_{L_1} 
                     \simeq &{1\over {\cal N}(L_1,L_2)(2\pi i)^{2}} 
    \oint   {dq_1} \, {dq_2} \,
                \exp\Biggl( \sum_{X_I\in S_U} {  q_1^{X_1}\, q_2^{X_2}
                \over 1- q_1^{X_1}\, q_2^{X_2}} \cr
           &+\log\left( \sum_{X_I\in S_U} { q_1^{X_1}\, q_2^{X_2}
                \over (1- q_1^{X_1}\, q_2^{X_2})^2 }\right) 
      - (L_1+1)\log q_1 - (L_2+1)\log q_2\Biggr). \cr 
}}
As in the 8D case, the average rank is approximately at $(L_1+L_2)/2$.
For $L_2=8$ and $(u_1,u_2)=(1,1)$ we have obtained the two curves in Figure 13.
Apparently, the SPA provides
a very good estimate. 
%begin figure
\fig{The average rank for coprime wrapping numbers with fixed  $L_2=8$ and
$(u_1,u_2)=(1,1)$. The dots  show the exact computer search result and
the solid  line the second order SPA.}{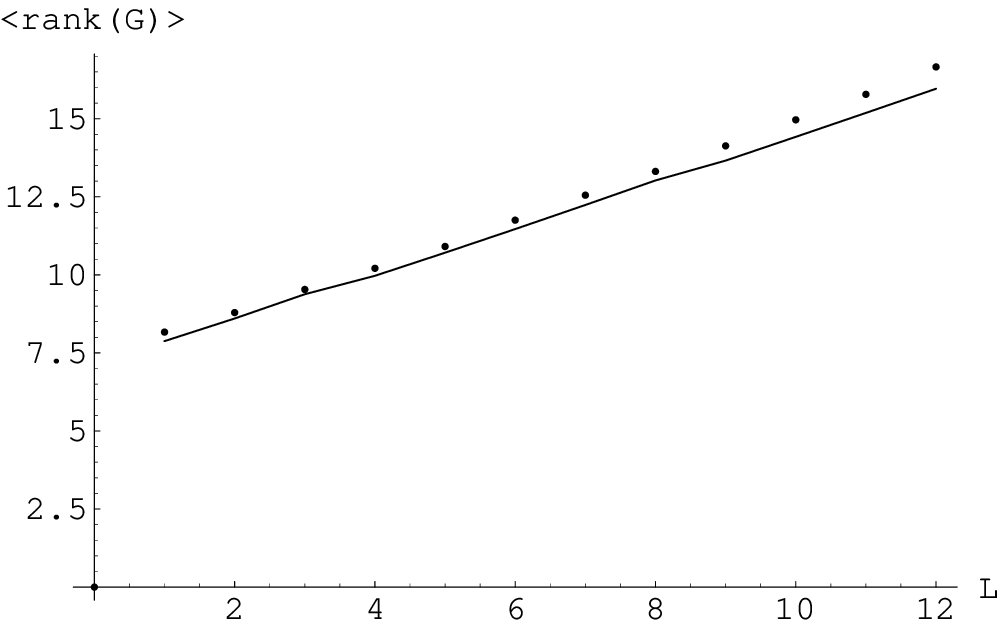}{10truecm}
%end figure

\subsec{The rank distribution}

The likelihood to find a gauge group of rank $r=\sum_a N_a$ can be determined
from the integral
\eqn\sixr{\eqalign{   P(r,L_I)
                     \simeq {1\over {\cal N}(L_1,L_2)\, (2\pi i)^{3}} 
    \oint   {dq_1} \, {dq_2}\, {dz} \,
                \exp\biggl(&\sum_{X_I\in S_U} {  z\, q_1^{X_1}\, q_2^{X_2}
                \over 1- z\, q_1^{X_1}\, q_2^{X_2}}- (L_1+1)\log q_1\cr
          & -(L_2+1)\log q_2- (r+1)\log z \biggr). \cr 
}}
In Figure 14 we show the distribution of the rank of the gauge group
for coprime wrapping numbers both for the exact computation
and the second order SPA. 
We see that the exact values scatter around the saddle point result. Since we
are not really at large values of the $L_I$, this is to be expected. 
The exact results still show number theoretic deviations from
the probably smooth asymptotic behaviour. 
%begin figure
\fig{The rank distribution for $L_1=10$, $L_2=8$ and $(u_1,u_2)=(1,1)$ and coprime
wrapping numbers. The dots are the exact results and the solid line the
SPA.}{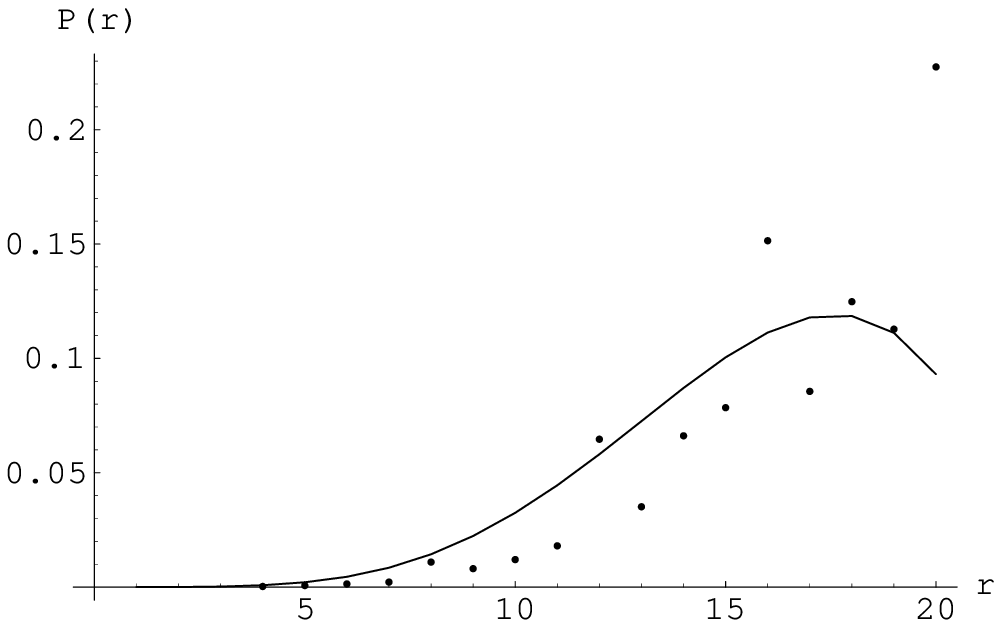}{10truecm}
%end figure

Figure 15 displays the saddle point result for the distribution of the rank of the gauge group
both for coprime and non-coprime wrapping numbers and $L_I=8$. 
For non-coprime wrapping numbers, the distribution 
again is  Gaussian, where the maximum lies approximately at $(L_1+L_2)/2$.
For coprime wrapping numbers the distribution also looks Gaussian, where
however the maximum has been shifted towards larger values of the rank.
This can be understood from the fact that for large values of the wrapping
numbers $(X_1,X_2)$ the coprime condition becomes harder to satisfy.
Large values of $(X_1,X_2)$ are therefore suppressed, which due to
the tadpole cancellation conditions means that smaller values of $N_a$ and
therefore of the rank appear less frequently. 
%begin figure
\fig{The rank distribution for $L_1=L_2=8$ and $(u_1,u_2)=(1,1)$. Dots: multiple wrapping, stars: coprime wrapping numbers.}{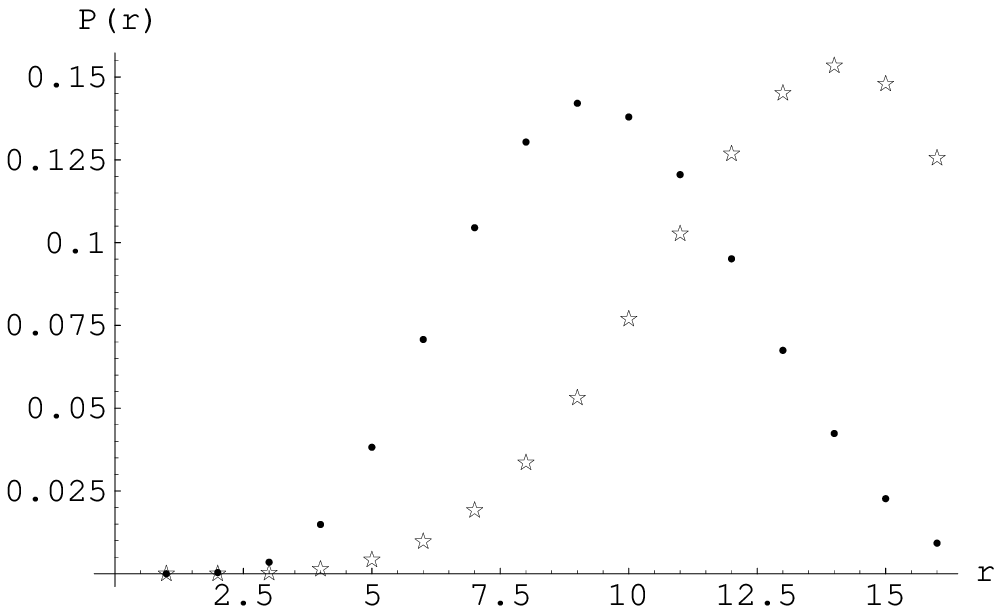}{10truecm}
%end figure
Let us also discuss what happens for other values 
of the complex structure moduli.
One example for coprime wrapping numbers  is shown in Figure 16, where we
compare for $L_1=L_2=16$ the distribution for $(u_1,u_2)=(1,1)$ and
$(u_1,u_2)=(20,1)$. Apparently, for large complex structures larger
values of the rank are preferred and the nice Gaussian shape of the
distribution is lost or rather the Gauss curve has been shifted
so much to large values of $r$ that only part of the left wing is visible. 
%begin figure
\fig{Comparison of  rank distributions for complex structures  $(u_1,u_2)=(1,1)$ (empty stars) and
$(u_1,u_2)=(20,1)$ (filled stars) in the case of coprime wrappings and $L_1=L_2=16$. }{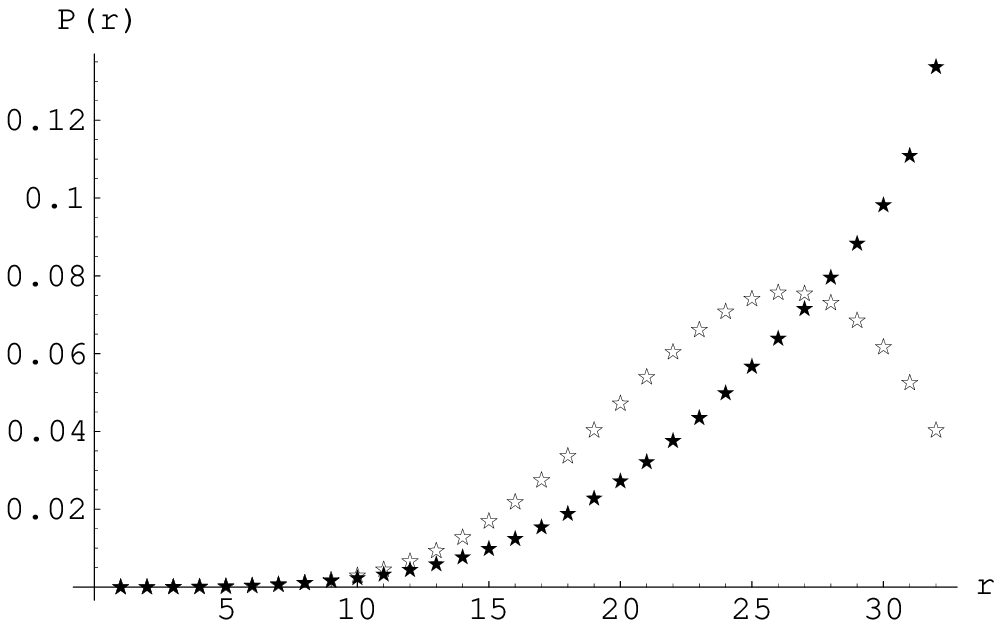}{10truecm}
%end figure

\subsec{Sum over complex structures}

So far we have just evaluated the various distributions for the
case that the $L_I$ and complex structures are fixed parameters.
Even though  for concrete applications  the $L_I$ are  indeed
fixed, the complex structures are not.
The values at which they are frozen by the supersymmetry constraints
depend on the concrete solution of  the tadpole cancellation conditions. 
In Figure 17 we show the allowed values of the coprime numbers $(u_1,u_2)$.
%begin figure
\fig{Allowed values of complex structures for $L_1=L_2=8$}{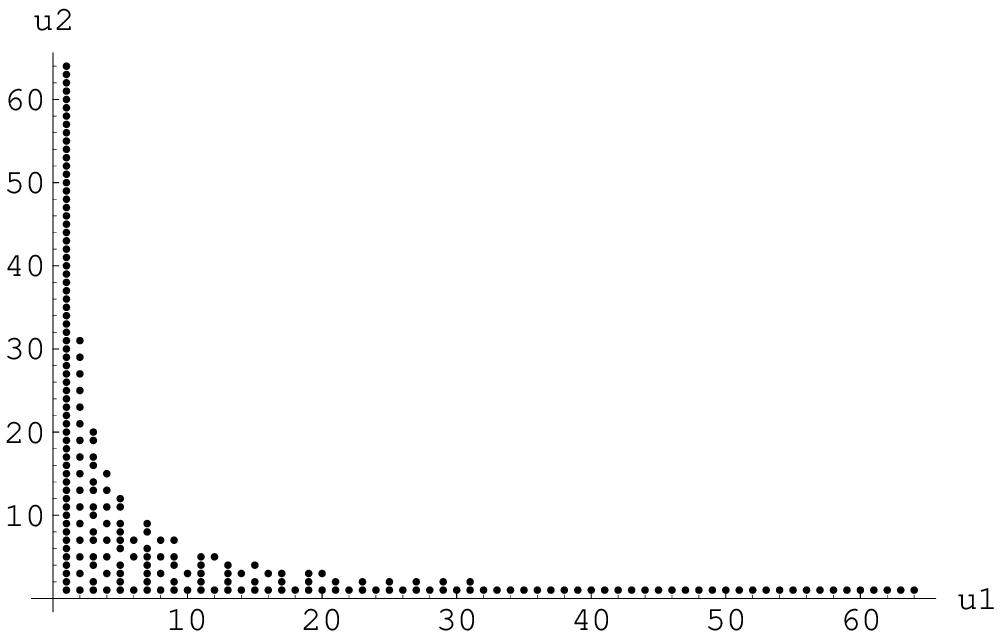}{10truecm}
%end figure
It is now an interesting question whether the summation over all
these complex structures destroys the qualitative features
we have seen for instance for the distribution of the rank of the
gauge group. 
It could very well be that any distinctive feature gets lost and
one eventually obtains  a uniform distribution.
We have carried out such an additional average for the case
$L_I=8$ with  the result shown in Figure 18.
Apparently, we still get shifted Gauss like curves, which shows
that the average over the complex structures is far from
leading to a uniform distribution.
%begin figure
\fig{Total distribution of the rank of the gauge group after adding up all
possible complex structures. We choose $L_1=L_2=8$. Dots: multiple wrapping, stars: coprime wrapping numbers.}{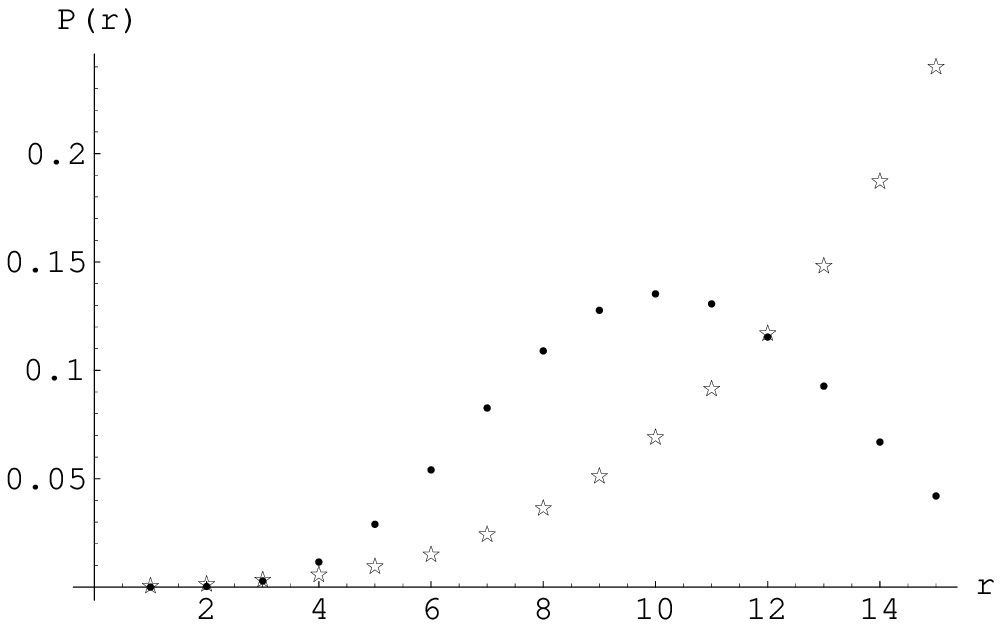}{10truecm}
%end figure

\subsec{The chirality distribution}

In order to proceed, one has to choose
a good measure for the chirality in the system. One possibility would be to
compute the distribution of intersection numbers $I_{ab}$ between different
stacks of branes. This is precisely what we will propose  in the 4D case,
but since the 6D case is not of phenomenological importance we choose
a simpler quantity involving only one set of branes like
$I_{a'a}=-8 X_{a,1}\, X_{a,2}$. This index  counts the net number of chiral
fermions  in the anti-symmetric plus  symmetric representation of the gauge group.
Since this index only depends on the wrapping numbers, the constraint
$X_{a,1}\, X_{a,2}=\chi$ can be easily implemented in the
saddle point integrals. The sum over one stack is restricted
to those wrapping numbers $S_{U,\chi}$ satisfying the constraint.

For the distribution of this intersection number in the ensemble of 6D
intersecting brane models with fixed complex structure and $L_I$ one
obtains 
\eqn\sixchiral{\eqalign{   P(\chi,L_I)
                     \simeq & {1\over {\cal N}(L_1,L_2)\, (2\pi i)^{2}} 
    \oint {dq_1} \, {dq_2}\, 
                \exp\biggl[\sum_{X_I\in S_U} { q_1^{X_1}\, q_2^{X_2}
                \over 1- q_1^{X_1}\, q_2^{X_2}}  \cr
                 &-\log\left(\sum_{X_I\in S_U} {   q_1^{X_1}\, q_2^{X_2}
                \over 1- q_1^{X_1}\, q_2^{X_2}}\right)+
              \log\left(\sum_{X_I\in S_{U,\chi}} {  q_1^{X_1}\, q_2^{X_2}
                \over 1- q_1^{X_1}\, q_2^{X_2}}\right) \cr
           & -(L_1+1) \log q_1- (L_2+1) \log q_2 \biggr]. \cr 
}}
Due to the supersymmetry constraint \roota, 
it is clear that for $(u_1,u_2)=(1,1)$  only
those $\chi$ appear which can be written as a square $\chi=\alpha^2$.
In Figure 19  we show the resulting distribution for $L_I=8$.
It is obvious that small values for the chirality index are favoured.
%begin figure
\fig{The chirality  distributions  for $L_I=8$ 
and $(u_1,u_2)=(1,1)$. Dots: multiple wrapping, stars: coprime wrapping numbers. }{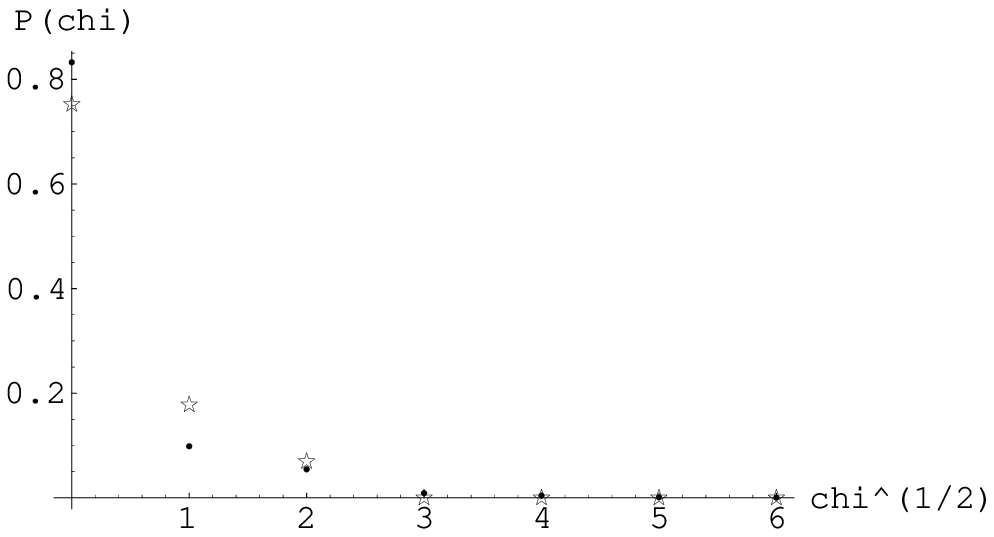}{10truecm}
%end figure
To get a better impression about the scaling, we show 
for the choice $L_I=16$ and  $(u_1,u_2)=(1,1)$ a logarithmic plot for non-coprime wrapping numbers in Figure 20.
From there we deduce that the scaling is roughly like $P(\chi)\simeq \exp(-\kappa \sqrt{\chi})$,
which could be expected as the wrapping numbers $X_{I,a}$ and the numbers of branes $N_a$
appear symmetrically in the tadpole cancellation conditions.
%begin figure
\fig{The chirality  distributions  for $L_I=16$, $(u_1,u_2)=(1,1)$ and
non-coprime wrapping numbers.}{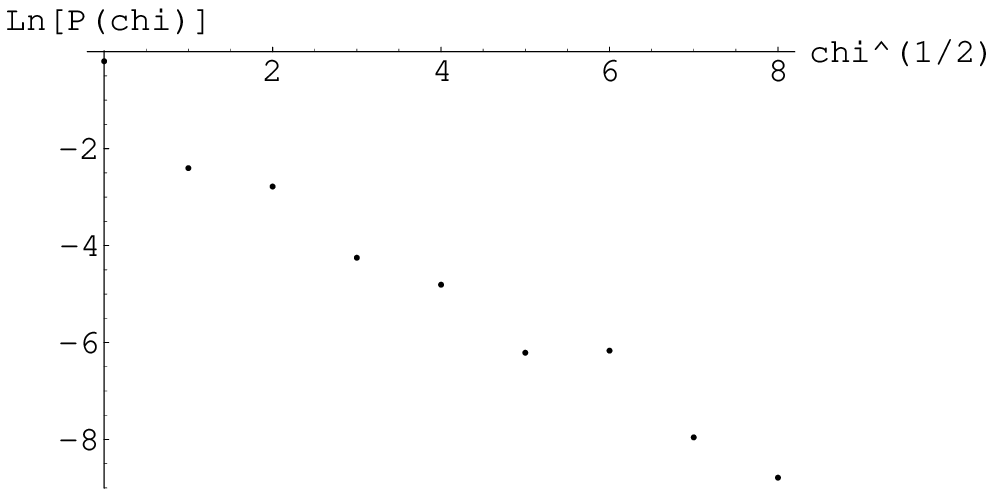}{10truecm}
%end figure

\subsec{Statistical correlations}

To conclude our discussion of the statistics of the 6D model, we compute the distribution
of the rank of the gauge group and the chirality index at the same time.
We are particularly interested in searching for a statistical correlation between these two quantities,
as expected from the tadpole cancellation conditions, which involve
the $N_a$ related to the rank and the $X_a$  related to $\chi$. 
The distribution is given by
\eqn\sixcorr{\eqalign{   P(\chi,r,L_I)
                     \simeq & {1\over {\cal N}(L_1,L_2)\, (2\pi i)^{3}} 
    \oint {dq_1} \, {dq_2}\, {dz} \, 
                \exp\biggl[\sum_{X_I\in S_U} { z\, q_1^{X_1}\, q_2^{X_2}
                \over 1- z\, q_1^{X_1}\, q_2^{X_2}}  \cr
                 &-\log\left(\sum_{X_I\in S_U} {   z\, q_1^{X_1}\, q_2^{X_2}
                \over 1- z\, q_1^{X_1}\, q_2^{X_2}}\right)+
              \log\left(\sum_{X_I\in S_{U,\chi}} {  z\, q_1^{X_1}\, 
                 q_2^{X_2}
                \over 1- z\, q_1^{X_1}\, q_2^{X_2}}\right) \cr
           & -(L_1+1)\log q_1- (L_2+1)\log q_2- (r+1)\log z  \biggr]. \cr 
}} 
Evaluating this threefold  integral via a numerical  saddle point approximation
results in the plot shown in figure 21.
%begin figure
\fig{Rank-chirality correlation for non-coprime wrapping numbers,
$L_I=8$ and $(u_1,u_2)=(1,1)$.}{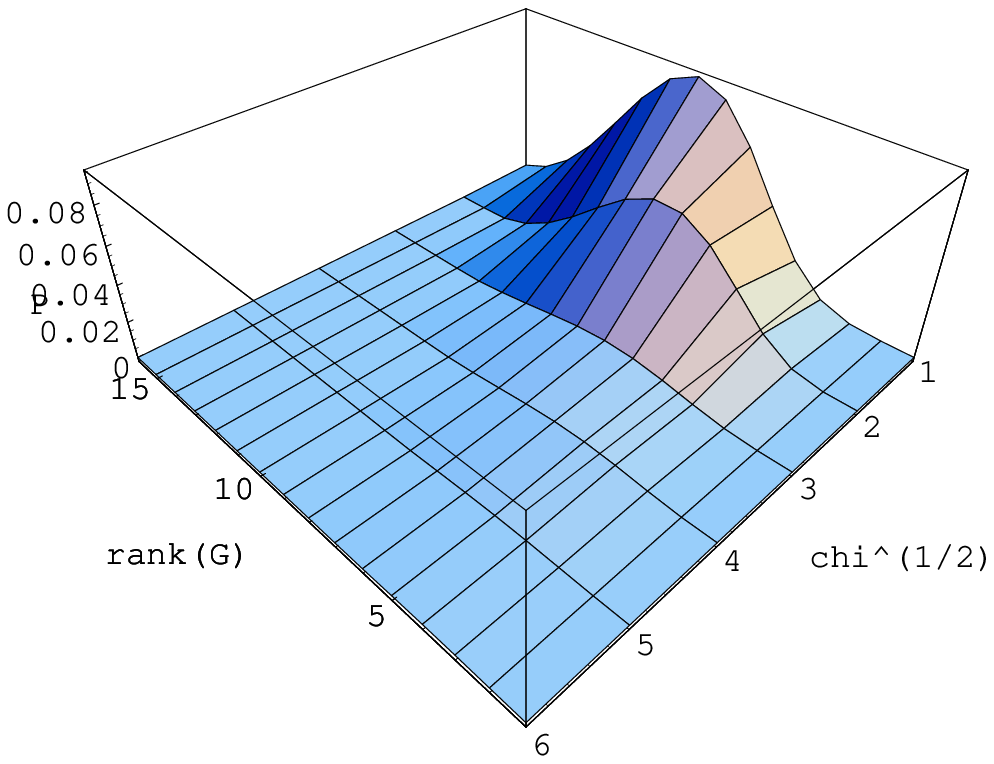}{10truecm}
%end figure
From this plot one can already guess that   the maximum of $r$ for fixed
$\chi$ depends on $\chi$. In fact it appears that for larger $\chi$ the
$r$ distribution moves towards smaller values of the rank.
Since the rank only depends on the various numbers of D-branes
and the chirality only on the wrapping numbers, the tadpole
cancellation conditions imply that these numbers are not independent.
In fact larger values of $\chi$ mean smaller values of the $N_a$.
What we see in the figure is the statistical manifestation
of this string theoretic constraint.
By looking at the plot from the another direction this correlation is 
even more obvious.
%begin figure
\fig{Rank-chirality correlation}{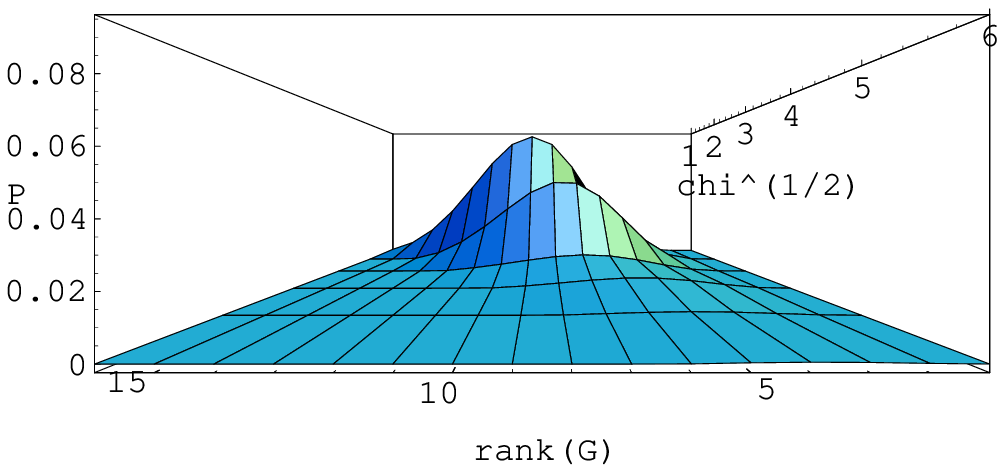}{10truecm} 
%end figure
We can also directly compute how for fixed values of the chirality $\chi$
the position of the maximum of $P(\chi,r)$ shifts towards smaller
values of $r$. The result is shown in Figure 23.
%begin figure
\fig{Shift of the maximum in $r$ for different values of $\chi$.}{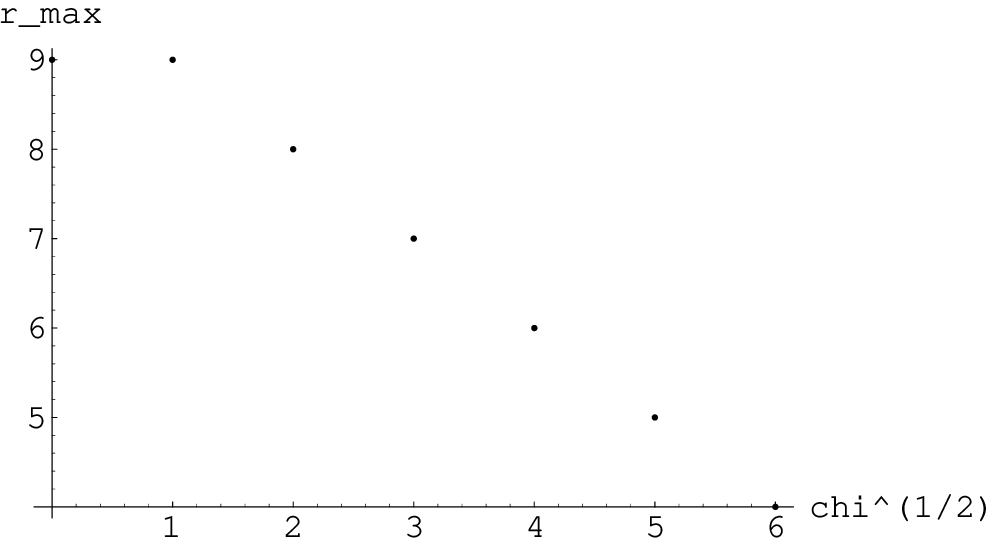}{10truecm}
%end figure
We conclude from this 6D intersecting D-brane model that the stringy tadpole
cancellation conditions imply a statistical correlation of the 
rank of the gauge group and its chiral matter content.

\newsec{Statistics of a 4D model}

Now we come to the physically most interesting case of a four 
dimensional model. 
In this section we first proceed completely analogously
to the higher  dimensional examples and  discuss the
various distributions for essentially fixed $L_I$.
However, it turned out that some computations already
become quite time consuming so that we postpone
their presentation to a future publication \bghlw.
Instead, in section 6 we will 
combine the gauge theory sector in the T-dual picture with a flux
compactification and perform the statistics taking
both sectors into account.

We now consider intersecting branes on the 
$T^6/\ZZ_2\times \ZZ_2$ orbifold without discrete torsion
\refs{\rcvetica}. As compared to the 6D case, this model shows
some new features, e.g. the possibility of negative contributions
to the tadpole cancellation conditions from supersymmetric 
branes. Therefore, it seems less obvious that also for
fixed complex structure only a finite number of solutions
exists. We will argue  in the next section that this 
is indeed the case. The reader who is not so interested in these
technical details may jump directly to section 5.2.

\subsec{Finiteness of solutions}

After recalling  the definitions of the model from section 2.2,
we would like to add that the coprime condition
on the wrapping numbers $(n_I,m_I)$ translates
into the following condition on the 3-cycle
wrapping numbers 
\eqn\coprimef{   \prod_{I=1}^3 {\rm gcd} (Y_0,X_I) = Y_0^2 .}
The proof for this can be found in Appendix B.
Now we would like to evaluate the various conditions the
wrapping numbers $(X_I,Y_I)$ have to satisfy in order to
see whether finitely many solutions exist.

First, we want to write the expressions more symmetrically
and transform $(X_I,Y_I)\to (-X_I,-Y_I)$ for $I=1,2,3$.
Moreover we define $U_0=1$, which allows us to write 
the supersymmetry conditions as
\eqn\susyca{\eqalign{ \sum_I Y_I {1\over U_I} &=0, \cr
                       \sum_I X_I {U_I} &>0, \cr}}
whereas the general consistency conditions are
\eqn\constca{\eqalign{ X_I\, Y_I &= X_J\, Y_J\quad {\rm for\ all}\ I,J, \cr
                       X_I\, X_J&=-Y_K\, Y_L, \cr
                       X_L\,(Y_L)^2&=-X_I\, X_J\, X_K, \quad\quad\quad  \cr
                       Y_L\,(X_L)^2&=-Y_I\, Y_J\, Y_K \quad {\rm for\ all}\ I\ne J\ne K\ne L\ne I. \cr}}
Multiplying each of the four tadpole equations
\eqn\tadab{\eqalign{    \sum_a N_a\, X_{a,I} &= L_I \cr }}
by $U_I$ and adding them all up yields together with \susyca\ the constraint
\eqn\constcb{   0<\sum_{I=0}^3 X_I\, U_I \le \sum_I L_I\, U_I .} 
Now let us distinguish the three cases that either one, two or four
of the $X_I$ are non-vanishing. The case that only one $X_I$ vanishes
is excluded by \constca.

For the case that only one $X_I$ is non-vanishing, the brane
lies precisely on top of one of the four orientifold planes
and supersymmetry imposes no constraint on the complex structures.

In the case that  two $X_I$ are non-vanishing, we are exactly in the
situation we discussed for the 6D model. Supersymmetry implies
that the $X_I>0$ so that the contributions to the
tadpole conditions are non-negative. In addition, one of
the three complex structures is fixed at a rational value, whereas
the other two remain as free parameters. 

Some new aspects appear for the case of all $X_I$ non-vanishing.
The constraints \constca\ imply immediately that an odd number
of the $X_I$ have to be negative. 

If three $X_i<0$ and, say, $X_A>0$, then the supersymmetry condition can be written as
\eqn\threenega{  \sum_I Y_I {1\over U_I}={Y_A\over U_A}\left( 1+\sum_i {X_A\,
U_A\over  X_i\, U_i}\right)=0, }
which implies $X_i\, U_i < -X_A\, U_A$ for all $i$. However, this is in contradiction
with the other supersymmetry condition 
\eqn\threenegb{  {X_A\, U_A}\left( 1+\sum_i {X_i\, U_i\over  X_A\, U_A}\right)>0, }
so that we conclude that no supersymmetric branes with three 
$X_i<0$ exist.

What remains is the case that three $X_i>0$ and one $X_A<0$. In this case
the same argument as above yields
$X_i\, U_i > -X_A\, U_A$ and 
does not give any contradiction.
Note that the supersymmetry condition \threenega\ allows
us to express $X_A$ in terms of the positive $X_i$
\eqn\threeposa{   X_A=-{1\over \sum_i {U_A\over U_i\, X_i} } .}
The upper bound in \constcb\ can be further evaluated as
\eqn\constcvc{  \sum_I L_I\, U_I \ge X_A\ U_A +\sum_{j} X_j\, U_j >
                   X_i\, U_i >0  } 
for any of the three possible $i$, where  we have used the relation
$X_i\, U_i > -X_A\, U_A$.

As in the 6D case, we can argue that if we take a sufficient
number of these branes, then the complex structures are fixed 
at rational values. So let us  write $U_I=u_{I,2}/u_{I,1}$ with
$u_{I,2},u_{I,1}\in \IN$ and relatively coprime.
Then the constraint \constcvc\ can be expressed as
\eqn\threeposc{ 1\le X_i\le {\sum_{P=0}^3   u_{P,2}u_{Q,1}u_{R,1}u_{S,1}
                 L_P  \over  u_{i,2}u_{J,1}u_{K,1}u_{L,1}} }
for $P\ne Q\ne R\ne S\ne P$ and $i\ne J\ne K\ne L\ne i$.
Thus, we conclude that for fixed complex structures only a finite
number of branes is admissible. 
It is much harder to see that for fixed $L_I$ also only a finite number
of complex structures allows any solution to the constraint \threeposc.
A computer analysis  however suggests that this is indeed the case.
If this is really true, then analogously to the 6D case, we would
argue that modulo moduli space identifications, for fixed
$L_I$ there exists only a finite number of solutions to the
tadpole cancellation conditions.

\subsec{Counting tadpole solutions}

Now we proceed completely along the lines of the higher dimensional
 examples and  compute
the total number of solutions to the tadpole cancellation conditions 
for fixed complex structures $U_1, U_2, U_3$.
Let us denote by $S_U$ the total set of the three classes of supersymmetric
branes described in the last section.
By now we are quite familiar with the integrals appearing so
that we can present them without too many comments,
\eqn\foura{\eqalign{   {\cal N}(L_I) 
                     \simeq {1\over (2\pi i)^{4}} 
    \oint   {dq_0} \, {dq_1}\, {dq_2}\, {dq_3} \,
                \exp\biggl(&\sum_{X_I\in S_U} {  \prod_I q_I^{X_I} 
                \over 1- \prod_I q_I^{X_I}} -
          \sum_I (L_I+1) \log q_I \biggr). \cr 
}}
The asymptotic growth of this expression can be deduced by the SPA
now in the four  variables $q_0,q_1,q_2,q_3$.
In Figure 24 we have shown the result of a numerical evaluation of this
saddle point.
\vskip 0.2cm
%begin figure
\fig{The total number of solutions for $L_1=L_2=L_3=8$. 
Dots: multiple wrapping, stars: coprime wrapping numbers.}{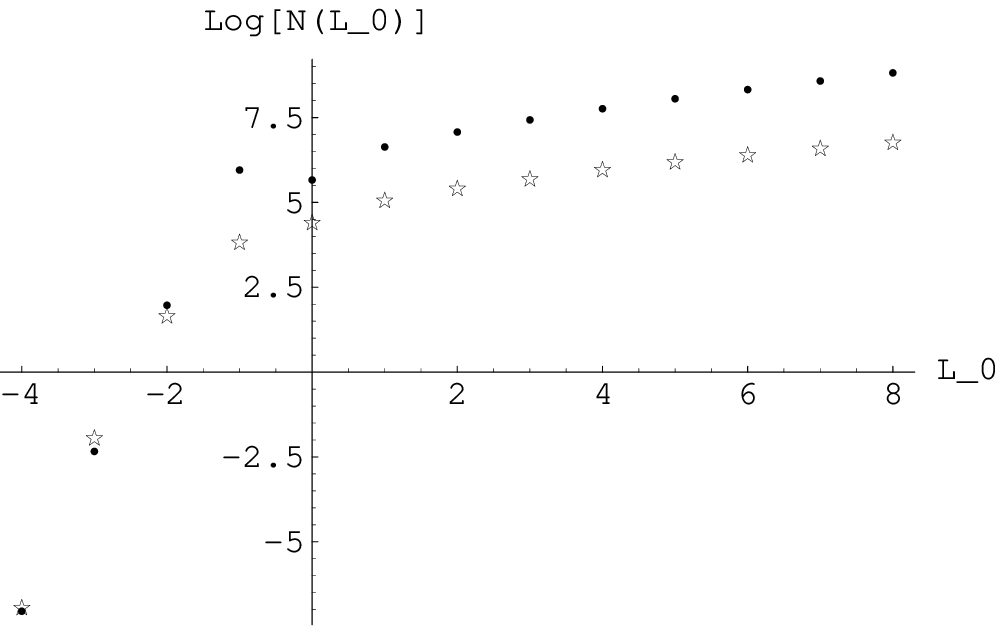}{10truecm}
%end figure
%If we exclude multiple wrapping of the branes we  get the result 
%shown in Figure ?
%\vskip 0.2cm
%begin figure
%\fig{The rank distribution}{plot1.eps}{10truecm}
%end figure
As compared to the 8D and 4D models, a new feature appears here which 
is rooted in the possibility to have supersymmetric
D-brane models even for the case $L_0<0$.
Looking at Figure 24, we observe that the curve  for $L_0\ge 0$
still scales like $\exp(\sqrt{L_1\log L_1})$,
even though one could have guessed that the new class
of branes with negative $X_I$ somehow changes
the behaviour.
However, for the case $L_0<0$ these latter branes become relevant, as
without at least one of them no solution to the tadpole cancellation
conditions can exist.  However, these branes have non-trivial winding
on all three $T^2$s, so that they are rather long leaving little room
for other branes. Therefore, we expect that for $L_0<0$
the number of solutions decreases dramatically as can be
nicely seen in Figure 24.

%It is also interesting to study the dependence of the number of solutions 
%on the complex structure moduli. In figure ? we show the dependence
%of the number of solutions on the allowed complex structures for fixed
%$L_1=8, L_2=8$,
%\vskip 0.2cm
%begin figure
%\fig{The rank distribution}{plot1.eps}{10truecm}
%end figure
%\meno
%Here we nicely see that the region of allowed complex up to moduli space
%identifications is bounded leading to finitely many solutions
%to the tadpole equations.
%{\bf comments }
%\meno

\subsec{Probability of $SU(M)$ gauge symmetry}

For fixed $L_I$ the probability to find an $SU(M)$ gauge factor
is given by
\eqn\fourb{\eqalign{   P(M,L_I) 
                     \simeq {1\over {\cal N}(L_I)(2\pi i)^{4}} 
    &\oint   \left(\prod_I {dq_I}\right)  \,
                \exp\Biggl(\sum_{X_I\in S_U} {  \prod_I q_I^{X_I}
                \over 1- \prod_I q_I^{X_I}} \cr
           &+\log\left( \sum_{X_I\in S_U} 
              \left(\prod_I q_I^{M X_I}\right)  \right) - \sum_I (L_I+1) \log q_I \Biggr). \cr 
}}
In Figure 25 we show the resulting distribution for $L_I=4\,\forall I$ and
realize that it also decreases exponentially for larger $M$.
\vskip 0.2cm
%begin figure
\fig{The probability for finding at least one SU(M) gauge factor for $L_I=4,\,\forall I$.
The result for non-coprime/coprime wrapping numbers is represented by the dots/stars.}{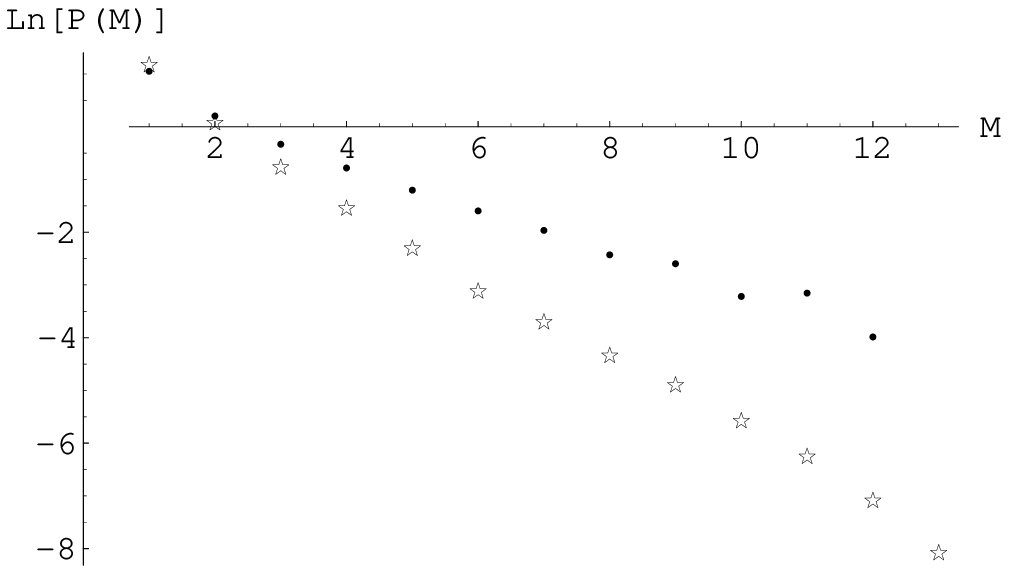}{10truecm}
%end figure
%For varying $L_I$ we find that a good approximation is given by
%\eqn\fourprob{ P(M, L_I)\simeq 
%\exp\left( -0.7\sqrt{\log L_1\over L_1}\, M  -0.7\sqrt{\log L_2\over L_2}\, M \right)
%} 
For $\sum_i M_i\ll \sum_I L_I$ we again find that the occurrence of a gauge factor
$\prod_i SU(M_i)$ is given by the product of the probabilities for the occurrence of 
each factor, $P(\vec M)=\prod_i P(M_i)$. 

For realising the Standard Model with  four intersecting branes, we learn
that an upper bound of their probability is $P_{SM}=P(3)P(2)P(1)P(1)$.
Of course, more conditions arise when we require
that the hypercharge $U(1)_Y$ remains massless after taking the
Green-Schwarz mechanism into account.

\subsec{The rank distribution}

The likelihood to find a gauge group of rank $r=\sum_a N_a$ can be determined
from the integral
\eqn\fourr{\eqalign{   P(r,L_I)
                     \simeq {1\over {\cal N}(L_I)\, (2\pi i)^{5}} 
    \oint   &{(\prod_I {dq_I}) \, {dz}} \,
                \exp\biggl(\sum_{X_I\in S_U} {  z\, \prod_I q_I^{X_I}
                \over 1- z\, \prod_I q_I^{X_I}} \cr
           & - \sum_I (L_I+1)\log q_I-
                (r+1) \log z \biggr). \cr 
}}
In Figure 26 we show the resulting rank distribution for $L_0=L_1=L_2=L_3=8$, 
$U_I=1$  and both non-coprime and
coprime wrapping numbers.
%begin figure
\fig{The rank distribution for $L_0=L_1=L_2=L_3=8$, $U_I=1$ and both non-coprime (dots) and
coprime (stars) wrapping numbers.}{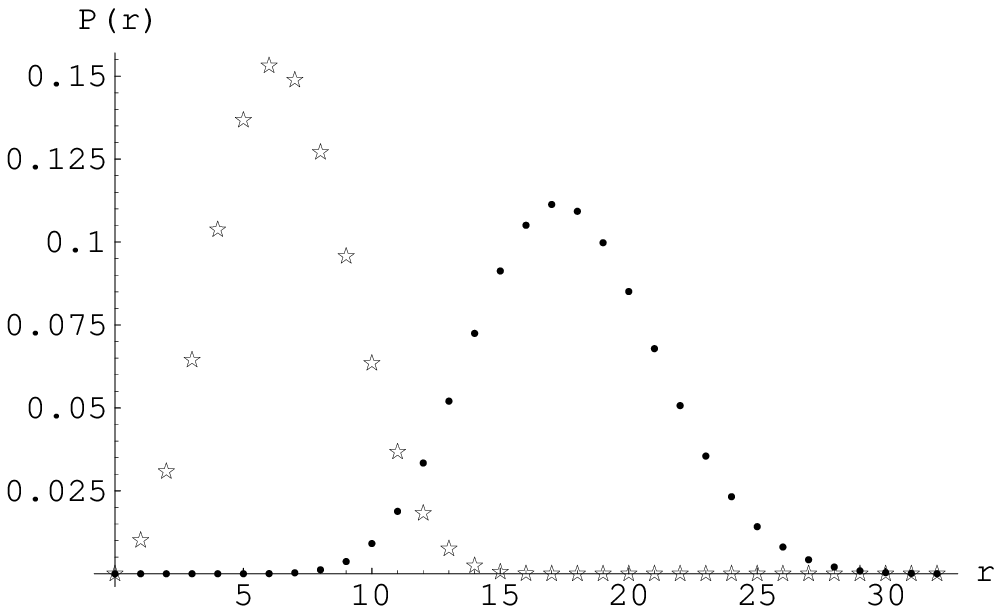}{10truecm}
%end figure
Since it is interesting to see what happens on the new branch $L_0<0$,
in Figure 27 we displayed the rank distributions for $L_1=L_2=L_3=8$ and
the choices $L_0=8$ and $L_0=-2$.
%begin figure
\fig{The rank distribution for $L_1=L_2=L_3=8$, $U_I=1$  and non-coprime wrapping numbers.
The dotted curve is for $L_0=-2$ and the one with triangles for $L_0=8$.}{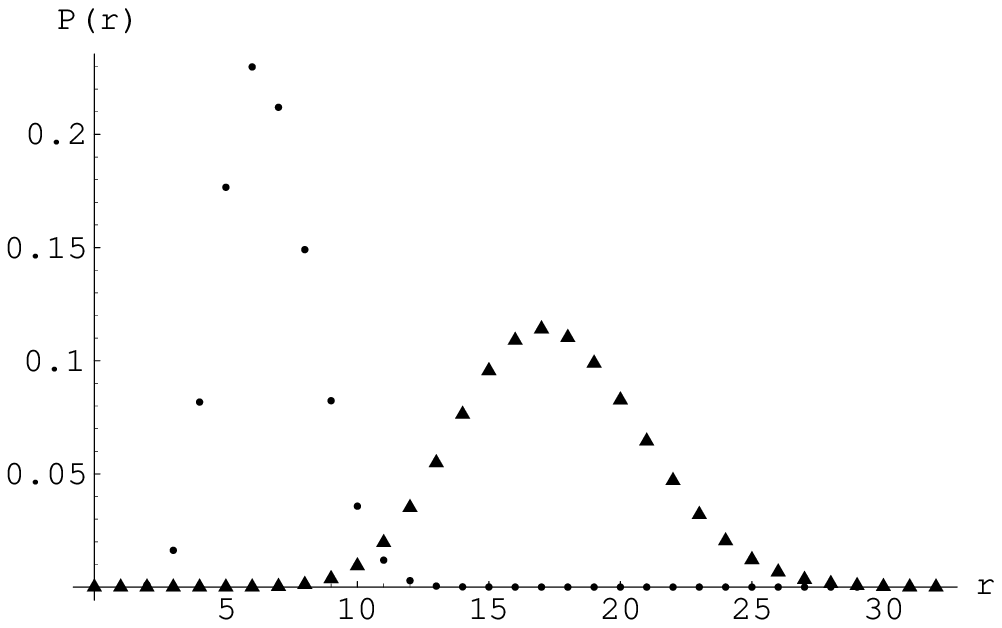}{10truecm}
%end figure}{10truecm}
%end figure}{10truecm}
%end figure}{10truecm}
%end figure
For $L_0\ge 0$ the distribution still has the familiar  Gaussian shape, 
where the maximum lies approximately at $(\sum_I L_I)/2$. 
For $L_0<0$ the shape of the curve is still Gaussian, but
the maximum is at a much smaller value than $r=(\sum_I L_I)/2$.
This can be understood by the fact that in this case
the special set of supersymmetric branes with some $X_a<0$ 
has to be present.
Since these branes are rather long, 
one expects the rank of the gauge group to be  reduced. 
%This can verified by plotting the rank distribution
%for $L_0=-2$ and $L_1=L_2=L_3=8$, as shown in Figure ?
%begin figure
%\fig{The rank distribution}{plot1.eps}{10truecm}
%end figure
%Instead of a broad maximum at $(\sum_I L_I)/2$ on obtains
%a narrow maximum for a small value of the rank.

\subsec{The chirality distribution}

Recall that in 6D we defined a measure for the chirality in the model
by the number of symmetric plus antisymmetric representations of a
gauge factor.
Here, to be closer to the question of the probability of a three generation
model, it is more appropriate to take the
intersection number $I_{ab}$ between two different stacks of D-branes.
More accurately we should take the number 
\eqn\chimeas{ \chi=  I_{a'b}-I_{ab}=2 \, \vec Y_a \vec X_b  ,}
which in our case of rectangular tori is always even, but
can become odd for tilted ones \rbkl.
For the distribution of this intersection number in the ensemble of 4D
intersecting brane models with fixed complex structure and $L_I$ one
obtains 
\eqn\fourchiral{\eqalign{   P(\chi,L_I)
                     \simeq & {1\over {\cal N}(L_I)\, (2\pi i)^{4}} 
    \oint (\prod_I {dq_I})\,  
                \exp\biggl[\sum_{X_I\in S_U} { \prod_I q_I^{X_I}
                \over 1- \prod_I q_I^{X_I} }  \cr
                 &-2\log\left(\sum_{X_I\in S_U} {  \prod_I q_I^{X_I}
                \over 1- \prod_I q_I^{X_I} }\right)
		+ \log\left(\sum_{X_{a,I}, X_{b,I}\in S_{U,\chi}} {  
                   \prod_I q_I^{X_{a,I}}
                \over 1- \prod_I q_I^{X_{a,I}}}{  
                   \prod_I q_I^{X_{b,I}}
                \over 1- \prod_I q_I^{X_{b,I}}}
\right) \cr
           &- \sum_I (L_I+1)\log q_I \biggr]. \cr 
}}
Our results for the chirality distribution displayed in Figure 28 clearly exhibit number 
theoretical effects in that the values for prime
$\chi \over 2$ tend to scatter around the approximate straight line formed by the other dots. 
Although much weaker, this effect can  already be observed in the corresponding 6D chirality 
distribution in Figure 20 and merely reflects the special factorization properties of 
prime numbers. 
\fig{The chirality distribution for $L_0 = L_1 =L_2 = L_3 = 8$, $U_I =1$
and multiple wrapping.}{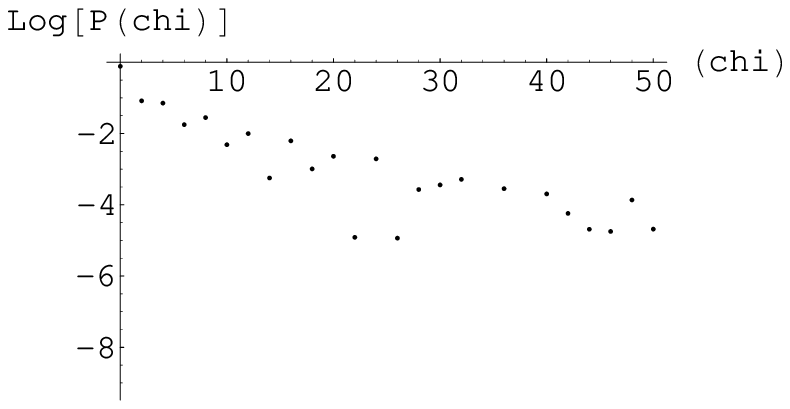}{10truecm}

%In figure [] we show the resulting distribution for $L_I=?$
%\vskip 0.2cm
%begin figure
%\fig{The rank distribution}{plot1.eps}{10truecm}
%end figure
%It is obvious that small values for the chirality index are favoured.
%the average values of $\chi$ is given by ???.

\subsec{Statistical correlations}

Finally, the statistical correlation between the rank of the gauge group and the chirality can be computed as

\eqn\fourcorr{\eqalign{   P(\chi,&r,L_I)
                     \simeq  {1\over {\cal N}(L_I)\, (2\pi i)^{5}} 
    \oint (\prod_I {dq_I})\, {dz}\, \, 
                \exp\biggl[\sum_{X_I\in S_U} { z\, \prod_I q_I^{X_I}
                \over 1- z\, \prod_I q_I^{X_I} }  \cr
                 &-2\log\left(\sum_{X_I\in S_U} {  z\, \prod_I q_I^{X_I}
                \over 1- z\, \prod_I q_I^{X_I} }\right)
		+ \log\left(\sum_{X_{a,I}, X_{b,I}\in S_{U,\chi}} {  
                   z\, \prod_I q_I^{X_{a,I}}
                \over 1- z\, \prod_I q_I^{X_{a,I}}}{  
                   z\, \prod_I q_I^{X_{b,I}}
                \over 1- z\, \prod_I q_I^{X_{b,I}}}
\right) \cr
           & -\sum_I (L_I+1)\log q_I - (r+1)\log z \biggr]. \cr 
}}
As can be seen from Figure 29, the maximum of the rank distribution takes smaller values for increasing chirality, in agreement  with the naive expectation from the tadpole cancellation 
conditions.
\fig{Rank-chirality correlation for $L_0=L_1=L_2=L_3=8,\, U_I=1$ and multiple wrapping.}{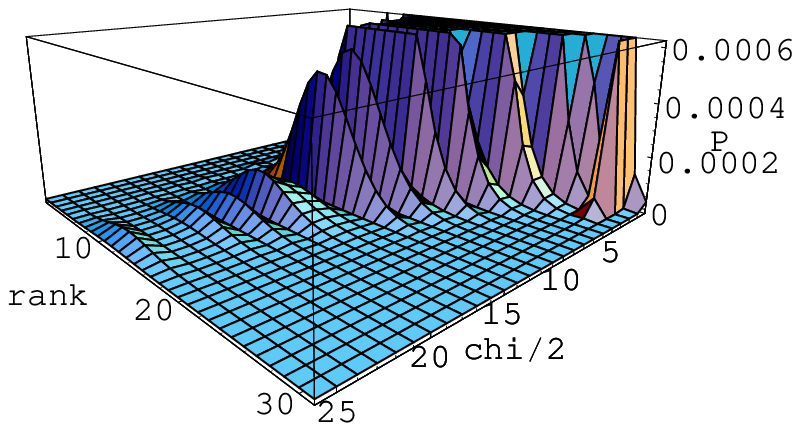}{10truecm}
%Note, that here we have already to numerically perform the saddle
%point approximation in six variables. But still the computer 
%needs only order of minutes for a complete run.
%For the correlation we obtain
%%begin figure
%\fig{Rank-chirality correlation}{corr6D.eps}{10truecm}
%%end figure
%From this plot one can already guess that  for instance the maximum of $r$ for fixed
%$\chi$ depends on $\chi$. In fact it looks like that for larger $\chi$ the
%$r$ distribution moves towards smaller values of the rank. 

%We can make this correlation even more evident by computing $P(\chi,r)-P(\chi)\, P(r)$
%which measures the deviation from statistical independence the rank
%of the gauge group and the chirality of its matter content.
%Figure ? features this correlation nicely, where for uncorrelated observables one would
%expect $P(\chi,r)-P(\chi)\, P(r)=0$.
%%begin figure
%\fig{Rank-chirality correlation}{corr6D.eps}{10truecm}
%%end figure
%We conclude that this 4D intersecting D-brane model shows that the stringy tadpole
%cancellation conditions imply a statistical correlation of the 
%rank of the gauge group and its chiral matter content. 

\newsec{Combination with flux compactification}

So far we have only discussed the case where the gauge theory sector of intersecting
branes is completely decoupled from the flux sector. That means that
in a concrete setting the non D-brane contributions $L_I$
to the tadpole cancellation conditions are fixed by the orientifold planes.
However, it is known that three-from fluxes in the Type IIB
T-dual models also give a positive contribution to the 4-form tadpole
condition. Therefore,  the effective non D-brane 
contribution to the 4-form tadpole is $L_0-N_{flux}$.
To avoid confusion in the following, please note that the former
complex structure moduli in the T-dual model become K\"ahler 
moduli and vice versa. 

In this section we would like to get a rough insight into
the statistical implications in the gauge theory sector
if we also take the degeneration of flux vacua into account.
In \AshokGK\ a formula for the number of flux vacua for given
$N_{flux} \leq L^*$ was derived, which  had the peculiar scaling
\eqn\scal{    {\cal N}\simeq   {(L^*)^K }, }
where $K$ is the number of three-cycles. This scaling
behaviour is a good approximation as long as $L^*\gg K$,
in which case the discrete sums could be estimated
by continuous integrals \refs{\AshokGK,\wgkt}. 
In the case we are discussing here, we have $L^*\le 8$ and
$K\simeq 10$, if we only allow for bulk fluxes, and
$K\simeq 100$, if we consider twisted 3-form fluxes, too.
Therefore, we are not really in 
the regime where we can trust the scaling \scal.
Nevertheless, as shown in \wgkt\ one can still  find
polynomial scalings for $L^*\sim K$, so  that we use  \scal\ as 
a rough estimate of what one can expect. 
In fact, we assume that the derivative of \scal\ gives the number
of flux vacua for fixed $N=L^*$.
Note that the aim here
is only to get a first glimpse of what can happen
when one combines the flux statistics  with the D-brane
statistics.
Note also that $N_{flux}$ is bounded
from above by the requirement that the conditions \constcvc\  admit
any non-trivial solution. Only branes with
all $X_I\ne 0$ can compensate a negative $L_0-N_{flux}$.

In the following, we leave the precise value of $K$ in \scal\
open and check how the various gauge sector distributions
change depending on the value of $K$.
To determine the statistics,  we now have to sum over all possible values of 
$N_{flux}$ and weight each term with the degeneracy \scal.
Since we expect the distributions not to depend strongly on the Type IIA
complex structure moduli (Type IIB K\"ahler moduli),
we are just choosing the specific value $U_I=1$ for simplicity.
For the rank distribution for instance we now get the following
expression
\eqn\fluxrank{  \overline{P}(r)={1\over N_{norm}} 
\sum_{N_{flux}=0}^{N_{flux}^{max}}
          {(N_{flux}+1)^K}\,\, {\cal N}(r;L_0-N_{flux},L_1,L_2,L_3), }
where ${\cal N}(r,L_I)$ is just the unnormalised part
of the distribution \fourr\ and $N_{norm}$ the new  
normalization constant. We introduced the factor $(N_{flux}+1)$ so that
$N_{flux}=0$ also contributes non-trivially.
In figure 30 we show the resulting distribution for the $\ZZ_2\times \ZZ_2$
orientifold. 
%begin figure
\fig{The rank distribution after averaging over flux vacua for $L_0=L_1=L_2=L_3=8$,
$U_I=1$ and $N_{flux}^{max}=11$.}{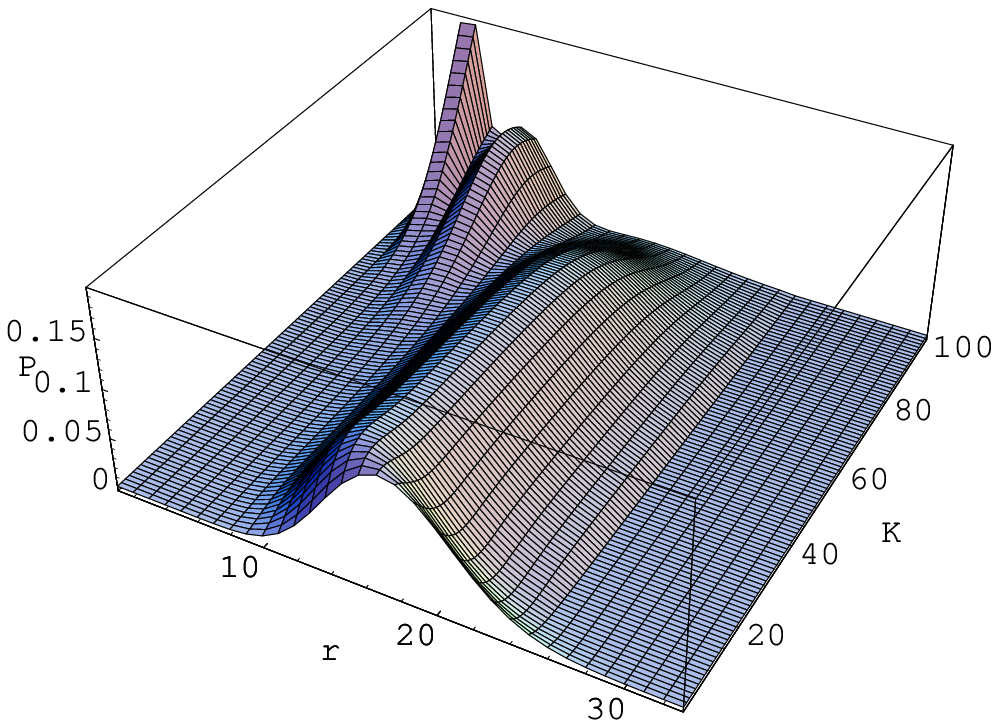}{15truecm}
%end figure
\noindent
From another angle the same distribution looks like shown
in Figure 31.
%begin figure
\fig{The rank distribution after averaging over flux vacua for $L_0=L_1=L_2=L_3=8$,
$U_I=1$ and $N_{flux}^{max}=11$.}{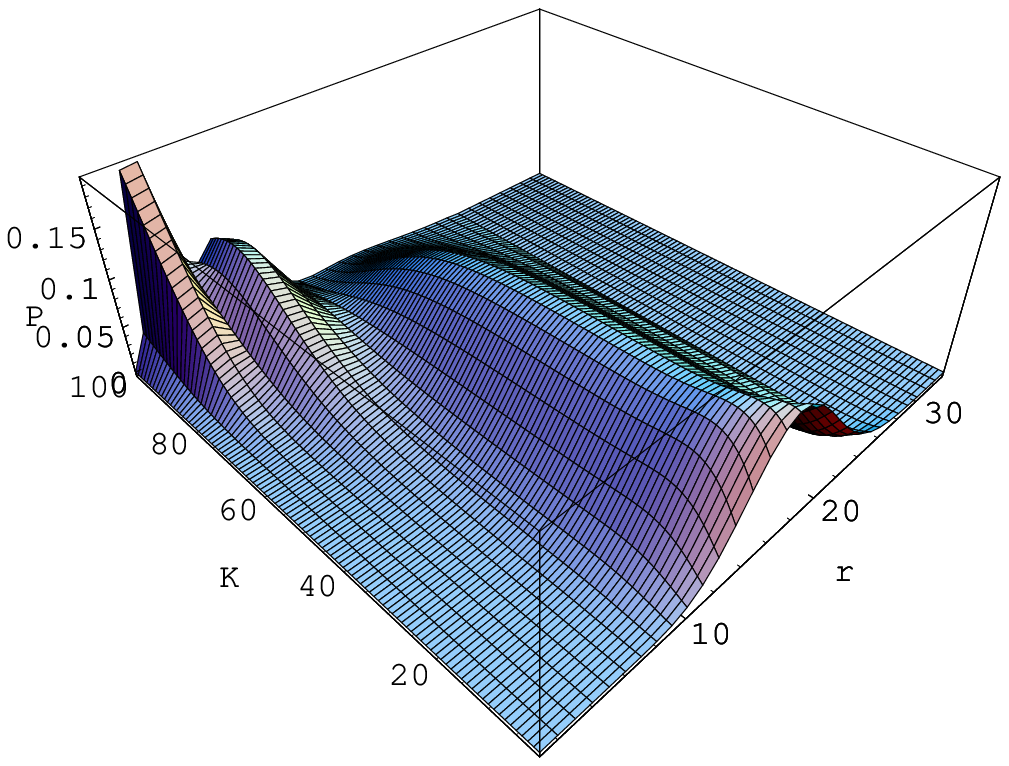}{15truecm}
%end figure
For a large range of values for $K$, one finds essentially the
same rank distribution as for fixed $L_0$, a Gauss curve with
a maximum at around $r_{max}\simeq 16$.
However for values $K>70$ new maxima appear for small values of $r$,
which come from the fact that in this case the large number of flux vacua
dominates the statistics and gives a larger weight to the actual
few brane solutions with $L_0<0$. Maybe it  has no actual significance,
but let us observe that the semi-realistic flux vacua constructed
in \MarchesanoXZ\  belong precisely to  this latter regime.

Naively, one could have expected that at least for certain values of
$K$ one obtains a uniform distribution for the rank of the gauge group,
but this does not seem to be the case. 
This first estimate of the influence of the flux sector on the
distribution of gauge theoretic observables seems to tell us that
even after averaging over the flux vacua one still gets non-trivial
distributions of the gauge theory observables.

%{\bf comments}

%Similarly we can also compute the new chirality-rank distribution,
%for which we obtain  figure ?
%begin figure
%\fig{Rank-chirality correlation}{corr6D.eps}{10truecm}
%end figure
%We conclude that in this model the most likely value for the rank of
%the gauge group having three generation is $r=?$.

Of course here we have just gained  a first glimpse and
it would be interesting to statistically analyse the effect on other
gauge theory observables and in particular on those, 
which depend on the Type IIB complex structure moduli and therefore
on the flux side of the model. More detailed physical quantities
like Yukawa couplings \yuka\ or soft supersymmetry breaking terms \LustDN\
fall into this category.

\newsec{Conclusions}

In this paper we have investigated the statistical behaviour
of solutions to the stringy tadpole cancellation conditions
as they appear for intersecting D-brane models.
Realizing that the problem of counting just solutions
is similar in spirit to the evaluation of the asymptotic growth
of the number of partitions, we have developed similar methods
for counting solutions to tadpole equations.
These methods are 
based on the saddle point approximation of the naturally
occurring integrals and could be generalised
to compute more sophisticated distributions of
physical quantities.
We have demonstrated for a simple enough 8D toy model
that this approximate method is indeed in good
agreement with an exact numerical computation
of all solutions. Even when the absolute values differ
by, say, one order of magnitude, the leading order 
saddle point approximation provides  already a very
good picture of the qualitative features of
the statistical distributions.

Encouraged by these observations we moved forward and  applied
similar methods to two concrete intersecting D-brane models in
6D and 4D. As a byproduct of our considerations
we proved and, respectively, gave very strong evidence that  
the number of solutions of the tadpole cancellation conditions
in the class of special Lagrangian cycles is finite, even
allowing varying complex structures.
We found that many of the qualitative features appearing in physically
more interesting 6D and 4D examples were already visible
in the 8D toy model. Intriguingly, in the former
models we observed a statistical correlation between
the number of families and the rank of the gauge group.
This confirms that string theory as a constrained system
can indeed give rise to non-trivial statistical correlations 
between physically relevant  and measurable quantities.
As was pointed out, such correlations might eventually
lead to a statistical falsification of string theory or
at least to statistical support for its relevance
in nature. Since we are just beginning to explore
the string theory landscape and develop mathematical methods for this purpose,
it is way too early to make any definite statement.

We are very optimistic that the general methods developed here can also
be applied to other statistical problems on the string theory
landscape like counting of other types of string theory vacua.
The moment one arrives at similar constraints for integer valued
quantities which admit  a large amount of solutions, these
methods should work.  We have in mind for instance counting
heterotic vacua like the subclass of toroidal orbifolds with
Wilson lines or the statistical analysis of Gepner model
orientifolds \refs{\gepner}. In the latter case
a systematic computer search for semi-realistic models
has been pioneered in \huiszoon.

Of course eventually one has to perform the statistics over
the ensemble of all string vacua and the various techniques developed
for certain aspects such as counting of flux vacua and
counting of tadpole solutions must be combined. We got a first
glimpse of what can happen in this case in section 6. 
Many ways to improve our understanding of the statistics on the string
theory landscape lie ahead and we are very curious to which
conclusions they will finally lead us. Can we ever falsify
string theory or provide at least strong statistical evidence for it or
do we have to face the unsatisfactory  conclusion that we
will never know?
But even if we can gain statistical evidence, can we then
move beyond this approach and really find at least some
of the realistic string vacua, which from the phenomenological
point of view would still be quite valuable, or do we have to
face the also depressing scenario  that - in the finite amount of time their
species exists - the best physicists can 
learn about the underlying structure of the universe, are
statistical correlations?

\centerline{{\bf Acknowledgements}}\pano

We would like to thank Prof. O. Forster and Peter Mayr for helpful discussion.
We are grateful to Andreas Wisskirchen and Thomas Hahn for technical support.

\vfill\eject
\noindent

\appendix{A}{Coprime wrapping numbers in 6D}

The quantities $X_I,Y_I$ are defined in \wrap\ and satisfy the relation
\consty. In this appendix we want to show how to recover the wrapping numbers
from given  $X_I,Y_I\in \ZZ$ and how to implement the coprime condition
which the wrapping numbers should fulfil.

Of course we have the problem that
not all chosen values for $X_I, Y_I \in \ZZ$ with $X_1\, X_2=Y_1\, Y_2$ correspond 
to configurations with  coprime wrapping numbers, e.g. take $X_1=-X_2=Y_2=-Y_1=2$. 
First we search for resolutions for non-coprime wrapping numbers 
$(n_I,m_I)=a_I\,  (p_I,q_I)$ with $p_I, q_I$ coprime.
Defining  $\alpha \equiv a_1 a_2$ we can write
\eqn\cond{\eqalign{
\alpha p_1 &=\gcd(X_1,Y_1), \cr
\alpha p_2 &=\gcd(X_1 ,Y_2 ),\cr
\alpha q_1 &=\gcd(X_2 ,Y_2),\cr
\alpha q_2 &=\gcd(X_2 ,Y_1, )
}}
and the coprime numbers are given by
\eqn\cone{\eqalign{
 p_1 &= \frac{X_1}{\gcd(X_1 ,Y_2 )} = \frac{Y_1}{\gcd(X_2 ,Y_1 )},\cr
p_2 &= \frac{X_1}{\gcd(X_1,Y_1)}= \frac{Y_2}{\gcd(X_2 ,Y_2) },\cr
q_1 &=\frac{X_2}{\gcd(X_2 ,Y_1 )}= \frac{Y_2}{\gcd(X_1 ,Y_2 )},\cr
q_2 &=\frac{X_2}{\gcd(X_2 ,Y_2)}= \frac{Y_1}{\gcd(X_1,Y_1) },
}}
while the global prefactor $\alpha$ can be written as
\eqn\conf{\eqalign{
\alpha &= \frac{\gcd (X_1,Y_2) \gcd (X_2,Y_2 )}{Y_2} \cr
 &= \frac{\gcd (X_1,Y_2) \gcd (X_2,Y_1 )}{X_1}  \cr
 &= \frac{\gcd (X_2,Y_1) \gcd (X_1,Y_1 )}{Y_1} \cr
 &= \frac{\gcd (X_2,Y_2) \gcd (X_2,Y_1 )}{X_2}. 
}}
From \cone\  and \conf\  one sees, that all $p_I, q_I, \alpha$ have exactly one expression in terms 
of $X_1, X_2, Y_2$. 
By construction, $p_I, q_I$ are integers, upon $Y_1 = \frac{X_1\, X_2}{Y_2} \in \ZZ$ 
one can also easily see that
the formulae above give an integer $\alpha$.

To summarize, given integers $X_1, X_2, Y_2$ with $\frac{X_1\,X_2}{Y_2} \in \ZZ$ 
can be rewritten in terms of 
coprime wrapping numbers $p_I, q_I$ and a global prefactor $\alpha$.
Now, the condition for no multiple wrappings reads  $\alpha=1$, from which 
we recover \gcda.

\appendix{B}{Coprime wrapping numbers in 4D}

With the definition from section 2, we first have to  recover 
the coprime integer wrapping numbers from $X_I, Y_I$.
We start as in the 6D case with non-coprime ones.
Now we search for resolutions for non-coprime wrapping numbers 
$(n_I,m_I)=a_I\,  (p_I,q_I)$ with $p_I, q_I$ coprime and
define $\alpha \equiv a_1 a_2 a_3$. Then we have
for $I\ne J\ne K\ne I, I,J,K \in \{1,2,3\}$
\eqn\gabd{\eqalign{
\alpha p_I\, p_J &=\gcd(X_0,{Y}_K),\cr
\alpha q_I\, q_J &=\gcd(Y_0,{X}_K),\cr
\alpha p_I\,  q_J &=\gcd({X}_I,{Y}_J),\cr
}}
and thereby
\eqn\gabe{\eqalign{
p_K &= \frac{X_0}{\gcd(X_0,{Y}_K)} = \frac{{Y}_J}{\gcd({X}_I,{Y}_J)}=
 \frac{{X}_K}{\gcd(Y_0,{X}_K)},  \cr
q_K &=  \frac{Y_0}{\gcd(Y_0,{X}_K)} = \frac{ {X}_I}{\gcd( {X}_I,{Y}_J)}=
 \frac{{Y}_K}{\gcd(X_0,{Y}_K) }, }}
and several relations of the form
\eqn\gabf{\eqalign{
\alpha p_K &=\frac{{X}_K}{q_I\, q_J} =\frac{X_0}{p_I\, p_J}=\frac{{Y}_I}{q_I\, p_J}, \cr
\alpha q_K &= \frac{{Y}_K}{p_I\, p_J}=\frac{Y_0}{q_I\, q_J}=\frac{{X}_I}{p_I\, q_J}, 
}}
which lead to the expression in terms of only $X_0,Y_0, {X}_K$
\eqn\gabg{
\alpha  = X_0\prod_{I=1}^3\frac{\gcd (Y_0,{X}_I)}{{X}_I} = 
 \frac{1}{(Y_0)^2} \prod_{I=1}^3 \gcd (Y_0,{X}_I) .}
The final condition for no multiple wrappings reads $\alpha=1$.
This gives equation \coprimef.

\appendix{C}{Summary of the computational technique}

In this section we would like to briefly  summarise the main computational
technique we used to determine the distribution
of various gauge theoretic observables in the ensemble
of intersecting D-brane models for fixed geometric background.

The first step is to determine all or at least a large,
preferably  representative subset of supersymmetric branes.
After solving the supersymmetry constraints, in all our examples 
this was given by a subset $S$ of
the naively allowed wrapping numbers $X_I$.
Then the total number of solutions to the tadpole cancellation
conditions
\eqn\summa{
       \sum_{a=1}^k  N_a\,  X_{a,I} = L_I 
}
with $I=1,\ldots,b_3/2$  is given by the expression
\eqn\summb{\eqalign{   {\cal N}(\vec L ) 
                     &\simeq {1\over (2\pi i)^{b_3\over 2}} 
    \oint \prod_I {dq_I \over q_I^{L_I+1} }\, 
                \exp\left(\sum_{X_I\in S} {\prod_I q_I^{X_I}
                \over 1-\prod_I q_I^{X_I}} \right).                     
}}
which can be  evaluated at leading order by a saddle point
approximation with 
\eqn\summc{
       f(\vec q)= \sum_{X_I\in S} {\prod_I q_I^{X_I}
                \over 1-\prod_I q_I^{X_I}} - \sum_I (L_I+1)\, \log q_I. 
}
The saddle point is determined by the condition $\nabla f(\vec q)|_{\vec q_0}=0$,
and the second order saddle point approximation reads
\eqn\summcc{ {\cal N}^{(2)}(\vec L)=  {1\over  \sqrt{2\pi}^{b_3 \over 2}}\, { e^{f(\vec q_{0})} 
         \over 
    \sqrt{ \det\left[ \left( {\partial^2 f\over \partial q_I\partial q_J}\right) 
        \right]_{q_0}}}.  }

An observable ${\cal O}$ in this example is given by a function 
${\cal O}({\bf N},{\bf  X}_I)$, where here ${\bf N}$ and ${\bf X}_I$
denote vectors with respect to the number of stacks. 
The expectation value of this observable in our ensemble
is defined as
\eqn\summd{\eqalign{   \langle {\cal O}({\bf N},{\bf  X}_I) \rangle(\vec L) 
                     \simeq {1\over {\cal N}(\vec L)\, (2\pi i)^{b_3\over 2}} 
    \oint \left( \prod_I {dq_I \over q_I^{L_I+1} }\right)\, 
              &\sum_{k=1}^\infty {1\over k!}\, 
                \sum_{N_1=1}^\infty \ldots
                  \sum_{N_k=1}^\infty \sum_{X_{1,I}\in S}\ldots
                \sum_{X_{k,I}\in S}\cr
                          & \left( \prod_I q_I^{{\bf N} {\bf X}_I}\right)
                       \,\,   {\cal O}({\bf N},{\bf X}_I)  \cr
}}
Depending on the actual form of ${\cal O}({\bf N},{\bf X}_I)$ these
sums can be further simplified and in the best case 
reduced to just a few sums over $S$. The final expression
is then to be evaluated using a saddle point approximation.

Similarly the distribution of ${\cal O}({\bf N},{\bf X}_I)=r$
is given by the expression
\eqn\summe{\eqalign{   P(r;\vec L)
                     \simeq {1\over {\cal N}(\vec L)\, (2\pi i)^{{b_3\over 2} +1}} 
    \oint \prod_I \left( {dq_I \over q_I^{L_I+1} }\right)\, {dz\over z^{r+1}}\, 
              &\sum_{k=1}^\infty {1\over k!}\, 
                \sum_{N_1=1}^\infty \ldots
                  \sum_{N_k=1}^\infty \sum_{X_{1,I}\in S}\ldots
                \sum_{X_{k,I}\in S}\cr
                          & \left( \prod_I q_I^{{\bf N} {\bf  X}_I}\right) \,
                         z^{{\cal O}({\bf N},{\bf X}_I )}. \cr
}}
Again simplifications can occur for certain choices of 
${\cal O}({\bf N},{\bf X}_I)$, which in effect
also simplify the saddle point approximation.

One can easily derive further generalisations of these expressions,
which in one way or the other will be appropriate to
study  similar statistical questions concerned with counting
solutions to discrete stringy consistency conditions. 

\listrefs

\bye
\end